\documentclass[11pt,a4paper,tightenlines,preprint,floatfix,nofootinbib,aps,prc,superscriptaddress,showkeys,showpacs]{revtex4}

\usepackage[dvips]{graphicx}
\usepackage{amsmath,amssymb,amsopn,bm,slashed}
\usepackage{color}
\usepackage{epstopdf}
\usepackage{grffile}

\usepackage[bookmarks,pdfhighlight=/O,colorlinks=false,pdfstartview=FitH]{hyperref}

\newcommand{\dd}{\text{d}}
\newcommand{\ee}{\text{e}}
\newcommand{\ii}{\text{i}}

\newcommand{\beq}{\begin{equation}}
\newcommand{\eeq}{\end{equation}}
\newcommand{\bce}{\begin{center}}
\newcommand{\ece}{\end{center}}

\newcommand{\GeV}{\ensuremath{\mathrm{GeV}}}

\begin{document}

\title{Asymptotic description of finite lifetime effects on the photon emission from a quark-gluon plasma}

\author{Frank Michler}
\email{michler@th.physik.uni-frankfurt.de}
\affiliation{Institut f{\"u}r Theoretische Physik, Goethe-Universit{\"a}t
Frankfurt, Max-von-Laue-Stra{\ss}e 1, D-60438 Frankfurt, Germany}

\author{Hendrik van Hees}
\email{hees@fias.uni-frankfurt.de}
\affiliation{Institut f{\"u}r Theoretische Physik, Goethe-Universit{\"a}t
Frankfurt, Max-von-Laue-Stra{\ss}e 1, D-60438 Frankfurt, Germany}
\affiliation{Frankfurt Institute for Advanced Studies (FIAS),
Ruth-Moufang-Stra{\ss}e 1, D-60438 Frankfurt, Germany}

\author{Dennis D. Dietrich}
\email{dietrich@th.physik.uni-frankfurt.de}
\affiliation{Institut f{\"u}r Theoretische Physik, Goethe-Universit{\"a}t
Frankfurt, Max-von-Laue-Stra{\ss}e 1, D-60438 Frankfurt, Germany}

\author{Carsten Greiner}
\email{carsten.greiner@th.physik.uni-frankfurt.de}
\affiliation{Institut f{\"u}r Theoretische Physik, Goethe-Universit{\"a}t
Frankfurt, Max-von-Laue-Stra{\ss}e 1, D-60438 Frankfurt, Germany}

\date{\today}

\begin{abstract}
 Direct photons play an important role as electromagnetic probes from the quark-gluon plasma (QGP) which occurs 
 during ultrarelativistic heavy-ion collisions. In this context, it is of particular 
 interest how the finite lifetime of the QGP affects the resulting photon production. Earlier investigations on 
 this question were accompanied by a divergent contribution from the vacuum polarization and by the remaining 
 contributions not being integrable in the ultraviolet (UV) domain. In this work, we provide a different 
 approach in which we do not consider the photon number density at finite times, but for free asymptotic 
 states obtained by switching the electromagnetic interaction according to the Gell-Mann and Low theorem. 
 This procedure eliminates a possible unphysical contribution from the vacuum polarization and, moreover, 
 renders the photon number density UV integrable. It is emphasized that the consideration of free asymptotic 
 states is, indeed, crucial to obtain such physically reasonable results.
\end{abstract}


\maketitle

\maketitle

\section{Introduction}  
\label{sec:intro}
Direct photons play an important role as electromagnetic probes for the quark-gluon plasma (QGP) which occurs 
during ultrarelativistic heavy-ion collisions \cite{Shuryak:1978ij,Shuryak:1977ut,Yag:2005,Muller:2006ee,Friman:2011zz}. Since photons 
interact only electromagnetically with the surrounding hadronic medium their mean free path is much larger than the spatial 
extension of the QGP. For that reason, they leave it almost undisturbed once they have been produced and therefore provide 
a direct insight into all stages of the collision. In this context, it is of particular interest how non-equilibrium 
effects such as the finite lifetime of the QGP affect the resulting photon emission.

Earlier investigations on this question \cite{Wang:2000pv,Wang:2001xh,Boyanovsky:2003qm} found that this finite lifetime gives rise to 
contributions from first-order QED processes, i.e., processes linear in the electromagnetic coupling constant, $\alpha_{e}$,  
which are kinematically forbidden in thermal equilibrium. Moreover, the photon spectrum resulting from these processes flattened 
into a power-law decay for photon energies $\omega_{\vec{k}}>1.5\; \GeV$ ($\omega_{\vec{k}}=|\vec{k}|$ with $\vec{k}$ denoting 
the three-momentum of the emitted photon), which would imply that in this domain the first-order contributions dominate over 
leading-order thermal contributions. The latter are linear in the electromagnetic coupling constant, $\alpha_{e}$, and the strong coupling constant, 
$\alpha_{s}$, in each case and thus of overall second order.

On the other hand, the investigations in \cite{Wang:2000pv,Wang:2001xh,Boyanovsky:2003qm} were accompanied by two serious 
artifacts. First, the photon number density contained a divergent contribution from the vacuum polarization for a given 
photon energy, $\omega_{\vec{k}}$. Moreover, the photon number density arising from the remaining contributions scaled as $1/\omega^{3}_{\vec{k}}$ 
in the ultraviolet (UV) domain. This implies that the total number density and the total energy density of the emitted photons are 
logarithmically and linearly divergent, respectively.

Recently, we have followed two other approaches in order to handle these problems in a consistent manner. In the first approach 
\cite{Michler:2009hi}, we have pursued a model description in which we have simulated the finite lifetime of the QGP by introducing 
time dependent quark/antiquark occupation numbers in the photon self-energy. This procedure allows for a consistent renormalization 
of the divergent contribution from the vacuum polarization. It does, however, not lead to a UV integrable photon number density for 
the general case.

At first we had suspected that this shortcoming results from a violation of the Ward-Takahashi identities within the model description 
\cite{Michler:2009hi}. For that reason, we have also pursued a second approach \cite{Michler:2012mg}, where we have modeled the creation 
of the QGP by a Yukawa-like source term in the QED-Lagrangian coupling the quarks and antiquarks to a purely time dependent, scalar 
background field. This effectively assigns the quarks and antiquarks a time dependent mass, which is consistent with the Ward-Takahashi 
identities. We have again restricted ourselves to first-order and thus purely non-equilibrium QED processes. These are kinematically possible 
in this case since the quarks and antiquarks obtain additional energy by the coupling to the time dependent background field. Similar investigations 
have been performed in \cite{Blaschke:2009uy,Blaschke:2010vs,Smolyansky:2010as,Blaschke:2011af,Blaschke:2011is} on electron-positron pair annihilation into a 
single photon in the presence of a strong laser field. There the preceding pair creation (and the subsequent annihilation) has been induced by a time dependent electromagnetic 
background field (see also \cite{Bethe01081934,Kluger:1992gb,Schmidt:1998vi,Pervushin:2006vh,Blaschke:2013ip}).

Another crucial difference to the approaches in \cite{Wang:2000pv,Wang:2001xh,Boyanovsky:2003qm,Michler:2009hi} has been the consideration 
of the photon number density not at finite times, but for free asymptotic states employing the standard Gell-Man and Low 
switching of the interaction Hamiltonian. Through this procedure, the photon number density is not plagued by the aforementioned unphysical 
contribution from the vacuum polarization anymore and, furthermore, has been rendered UV integrable for suitable mass parameterizations, $m(t)$. In 
particular, our investigations have shown that the photon number density indeed has to be considered for free asymptotic states in order to 
obtain such physically reasonable results. In this context, we have seen that a consistent definition of the photon number density is actually 
only possible for such free asymptotic states, whereas a similar interpretation of the respective expression is usually not justified at finite times, $t$. 
Such a conceptual problem also occurs if the electromagnetic interaction is only switched on from $t\rightarrow-\infty$ 
but not off again for $t\rightarrow\infty$, which has been suggested in \cite{Serreau:2003wr} in order to implement initial correlations at some 
$t=t_{0}$ developing from an uncorrelated initial state at $t\rightarrow-\infty$. Hence, the results from \cite{Michler:2012mg} raise the question whether 
the artifacts encountered in \cite{Wang:2000pv,Wang:2001xh,Boyanovsky:2003qm,Michler:2009hi} result from an inconsistent definition of the `photon number density' 
at finite times and whether they are removed if this quantity is considered for free asymptotic states instead.

Accordingly, in this work we revisit the previous approach \cite{Michler:2009hi}. This means that we again simulate the time-evolution of the 
QGP during a heavy-ion collision by introducing strongly time dependent quark/antiquark occupation numbers in the photon self-energy, but we 
consider photon number density not at finite times, but for free asymptotic states. Hence, we adhere to our principle approach from
\cite{Michler:2012mg} but consider an alternative description for our time dependent emitting system. We shall demonstrate that in direct analogy to 
\cite{Michler:2012mg}, this procedure again eliminates a potential unphysical contribution from the vacuum polarization. Moreover, it
leads to a UV integrable photon number density if the time evolution of the quark/antiquark occupation numbers in the photon self-energy is 
described in a physically reasonable manner, i.e., if it is taken into account that these occupation numbers are populated over a finite interval of 
time. In this context, we emphasize again that considering the photon number density for free asymptotic states is, indeed, crucial to obtain such physically 
reasonable results and that the artifacts encountered in \cite{Wang:2000pv,Wang:2001xh,Boyanovsky:2003qm} and still partly in \cite{Michler:2009hi} 
would reappear if this quantity were considered at finite times.

This paper is organized as follows: In section \ref{sec:model}, we provide a detailed description of our (revised) model approach on first-order 
photon production from a QGP. In particular, we demonstrate how we simulate the time evolution of the QGP by introducing fastly populating, time dependent 
quark/antiquark occupation numbers in the photon self-energy and how our asymptotic description eliminates a possible unphysical 
contribution from the vacuum polarization. After that, we present our numerical investigations in section \ref{sec:numint}. We show that in this present setting, 
our description also leads to a UV integrable photon number density. There we also provide detailed considerations on the dependence of the photon number density 
on the time scale, $\tau$, over which the quark/antiquark occupation numbers are assumed to build up. Then we compare our results to leading-order thermal photon 
emission in section \ref{sec:thermcomp}. In section \ref{sec:aspt}, we again highlight the necessity to consider the photon number density for free asymptotic 
states before we finish with a summary and an outlook to future investigations in section \ref{sec:conclusions}. Technical details are given in appendix 
\ref{sec:appa}.

\section{Asymptotic photon number density}
\label{sec:model}
Before we start with our numerical investigations, we provide a more extensive description of our model approach than given in \cite{Michler:2009hi}. 
The starting point is the photon number density for a homogeneous, but non-stationary emitting system of deconfined quarks and antiquarks. At first 
order in $\alpha_{e}$, this quantity is given by
\begin{equation}
 \label{eq:2:photnum}
 2\omega_{\vec{k}}\frac{\dd^{6}n_{\gamma}(t)}{\dd^{3}x\dd^{3}k} = 
   \frac{1}{(2\pi)^{3}}\int_{-\infty}^{t}\dd t_{1}\int_{-\infty}^{t}\dd t_{2}\ii\Pi^{<}_{\text{T}}(\vec{k},t_{1},t_{2})
   \ee^{\ii\omega_{\vec{k}}(t_{1}-t_{2})} \ .
\end{equation}
Here $\ii\Pi^{<}_{\text{T}}(\vec{k},t_{1},t_{2})$ denotes the transverse part of the photon self-energy, i.e., 
\begin{equation}
 \label{eq:2:contraction}
 \ii\Pi^{<}_{\text{T}}(\vec{k},t_{1},t_{2}) = \gamma^{\mu\nu}(\vec{k})\ii\Pi^{<}_{\nu\mu}(\vec{k},t_{1},t_{2}) \ .
\end{equation}
$\gamma^{\mu\nu}(\vec{k})$ is the photon tensor reading
\begin{equation}
 \label{eq:2:polten}
 \gamma^{\mu\nu}(\vec{k}) = \sum_{\lambda=\perp}\epsilon^{\mu,*}(\vec{k},\lambda)\epsilon^{\nu}(\vec{k},\lambda)
                          = \begin{cases}
                             -g^{\mu\nu}-\frac{k^{\mu}k^{\nu}}{\omega^{2}_{\vec{k}}} & \ , \quad\mbox{for}\quad\mu,\nu\in\left\lbrace1,2,3\right\rbrace \\
                              0                                                      & \ , \quad\mbox{otherwise}
                            \end{cases} \ ,
\end{equation}
where the sum runs over all physical (transverse) polarizations. Moreover, we have introduced the four vector $k^{\mu}=(\omega_{\vec{k}},\vec{k})$.
The photon self-energy, $\ii\Pi^{<}_{\mu\nu}(\vec{k},t_{1},t_{2})$, in turn is given by the thermal one-loop approximation
\begin{equation}
 \label{eq:2:pse_loop}
 \ii\Pi^{<}_{\mu\nu}(\vec{k},t_{1},t_{2}) = e^{2}\int\frac{\dd^{3}p}{(2\pi)^{3}}\mbox{Tr}
                                                  \left\lbrace 
                                                   \gamma_{\mu}S^{<}_{\text{F}}(\vec{q},t_{1},t_{2})
                                                   \gamma_{\nu}S^{>}_{\text{F}}(\vec{p},t_{2},t_{1})
                                                  \right\rbrace \ , 
\end{equation}
where $e$ denotes the electromagnetic coupling and $\vec{q}=\vec{p}+\vec{k}$. In thermal equilibrium, the fermion propagators entering 
(\ref{eq:2:pse_loop}) read
\begin{subequations}
 \label{eq:2:propagators}
 \begin{eqnarray}
  S^{<}_{\text{F}}(\vec{q},t_{1},t_{2}) & = & S^{<}_{\text{Q}}(\vec{q},t_{1},t_{2})+S^{<}_{\text{AQ}}(\vec{q},t_{1},t_{2}) \ , \\
  S^{>}_{\text{F}}(\vec{p},t_{1},t_{2}) & = & S^{>}_{\text{Q}}(\vec{p},t_{1},t_{2})+S^{>}_{\text{AQ}}(\vec{p},t_{1},t_{2}) \ ,
 \end{eqnarray}
\end{subequations}
with the quark (Q) and antiquark (AQ) components
\begin{subequations}
 \label{eq:2:prop_split}
 \begin{eqnarray}
  S^{<}_{\text{Q}}(\vec{q},t_{1},t_{2})  & = & \ii n_{\text{F}}(q_{0})\frac{\slashed{q}+m}{2q_{0}}\cdot\ee^{-\ii q_{0}(t_{1}-t_{2})} \label{eq:2:prop_split_a} \ , \\
  S^{<}_{\text{AQ}}(\vec{q},t_{1},t_{2}) & = & \ii\left[1-n_{\text{F}}(q_{0})\right]\frac{\slashed{\bar{q}}-m}{2q_{0}}\cdot\ee^{\ii q_{0}(t_{1}-t_{2})} \label{eq:2:prop_split_b} \ , \\
  S^{>}_{\text{Q}}(\vec{p},t_{1},t_{2})  & = & -\ii\left[1-n_{\text{F}}(p_{0})\right]\frac{\slashed{p}+m}{2p_{0}} \cdot\ee^{-\ii p_{0}(t_{1}-t_{2})} \label{eq:2:prop_split_c} \ , \\
  S^{>}_{\text{AQ}}(\vec{p},t_{1},t_{2}) & = & -\ii n_{\text{F}}(p_{0})\frac{\slashed{\bar{p}}-m}{2p_{0}}\cdot\ee^{\ii p_{0}(t_{1}-t_{2})} \label{eq:2:prop_split_d} \ .
 \end{eqnarray}
\end{subequations}
Here $n_{\text{F}}(E)$ is the Fermi-Dirac distribution function 
\begin{equation}
 \label{eq:2:fermi-dirac}
 n_{\text{F}}(E)=\frac{1}{1+\ee^{\beta E}} \ ,
\end{equation}
with $\beta=1/T$ and $T$ denoting the temperature of the system. Moreover, we have introduced the four-vector notations $p^{\mu}=(E_{\vec{p}},\vec{p})$ 
and $\bar{p}^{\mu}=(E_{\vec{p}},-\vec{p})$. Here $E_{\vec{p}}\equiv\sqrt{p^2+m^{2}}$ is the free relativistic quark/antiquark energy with 
$p$ and $m$ describing the absolute value of the three-momentum, $\vec{p}$, and the quark/antiquark mass, respectively. 

It follows from (\ref{eq:2:prop_split}) that expression (\ref{eq:2:pse_loop}) contains the contributions from the four first-order QED 
processes. These processes are (one-body) quark Bremsstrahlung (QBS), (one-body) antiquark Bremsstrahlung (ABS), quark-antiquark pair annihilation into 
a single photon (ANH), and the spontaneous creation of a quark-antiquark pair together with a photon out of the vacuum (PAC). Hence it is convenient 
to split up (\ref{eq:2:pse_loop}) accordingly, i.e., 
\begin{equation}
 \ii\Pi^{<}_{\mu\nu}(\vec{k},t_{1},t_{2}) = \ii\Pi^{\text{QBS}}_{\mu\nu}(\vec{k},t_{1},t_{2})+\ii\Pi^{\text{ABS}}_{\mu\nu}(\vec{k},t_{1},t_{2})+
                                            \ii\Pi^{\text{ANH}}_{\mu\nu}(\vec{k},t_{1},t_{2})+\ii\Pi^{\text{PAC}}_{\mu\nu}(\vec{k},t_{1},t_{2}) \ ,
\end{equation}
with the particular contributions given by
\begin{subequations}
 \label{eq:2:pse_split}
  \begin{eqnarray}
   \ii\Pi^{\text{QBS}}_{\mu\nu}(\vec{k},t_{1},t_{2}) & = & e^{2}\int\frac{\dd^{3}p}{(2\pi)^{3}}\text{Tr}\left\lbrace\gamma_{\mu}
                                                           S^{<}_{\text{Q}}(\vec{q},t_{1},t_{2})\gamma_{\nu}S^{>}_{\text{Q}}(\vec{p},t_{2},t_{1})\right\rbrace
                                                           \label{eq:2:pse_split_qbs} \ ,  \\
   \ii\Pi^{\text{ABS}}_{\mu\nu}(\vec{k},t_{1},t_{2}) & = & e^{2}\int\frac{\dd^{3}p}{(2\pi)^{3}}\text{Tr}\left\lbrace\gamma_{\mu}
                                                           S^{<}_{\text{AQ}}(\vec{q},t_{1},t_{2})\gamma_{\nu}S^{>}_{\text{AQ}}(\vec{p},t_{2},t_{1})\right\rbrace
                                                          \label{eq:2:pse_split_abs} \ , \\
   \ii\Pi^{\text{ANH}}_{\mu\nu}(\vec{k},t_{1},t_{2}) & = & e^{2}\int\frac{\dd^{3}p}{(2\pi)^{3}}\text{Tr}\left\lbrace\gamma_{\mu}
                                                           S^{<}_{\text{Q}}(\vec{q},t_{1},t_{2})\gamma_{\nu}S^{>}_{\text{AQ}}(\vec{p},t_{2},t_{1})\right\rbrace
                                                           \label{eq:2:pse_split_anh} \ , \\
   \ii\Pi^{\text{PAC}}_{\mu\nu}(\vec{k},t_{1},t_{2}) & = & e^{2}\int\frac{\dd^{3}p}{(2\pi)^{3}}\text{Tr}\left\lbrace\gamma_{\mu}
                                                           S^{<}_{\text{AQ}}(\vec{q},t_{1},t_{2})\gamma_{\nu}S^{>}_{\text{Q}}(\vec{p},t_{2},t_{1})\right\rbrace
                                                           \label{eq:2:pse_split_pac} \ .
  \end{eqnarray}
\end{subequations}
It follows from (\ref{eq:2:prop_split_a})-(\ref{eq:2:prop_split_d}) that the contraction with $\gamma^{\mu\nu}(\vec{k})$ yields
\begin{subequations}
 \label{eq:2:pse_split_trans}
  \begin{align}
   \ii\Pi^{\text{QBS}}_{\text{T}}(\vec{k},t_{1},t_{2}) = 2e^{2}\int\frac{\dd^{3}p}{(2\pi)^{3}} 
          & \left\lbrace1-\frac{px(px+\omega_{\vec{k}})+m^{2}}{p_{0}q_{0}}\right\rbrace
            n_{\text{F}}(q_{0})\left[1-n_{\text{F}}(p_{0})\right] \nonumber \label{eq:2:pse_split_trans_qbs} \\
   \times & \text{ }\ee^{-\ii(q_{0}-p_{0})(t_{1}-t_{2})} \ , \\
   \ii\Pi^{\text{ABS}}_{\text{T}}(\vec{k},t_{1},t_{2}) = 2e^{2}\int\frac{\dd^{3}p}{(2\pi)^{3}} 
          & \left\lbrace1-\frac{px(px+\omega_{\vec{k}})+m^{2}}{p_{0}q_{0}}\right\rbrace
            n_{\text{F}}(p_{0})\left[1-n_{\text{F}}(q_{0})\right] \nonumber \label{eq:2:pse_split_trans_abs} \\
   \times & \text{ }\ee^{\ii(q_{0}-p_{0})(t_{1}-t_{2})} \ , \\
   \ii\Pi^{\text{ANH}}_{\text{T}}(\vec{k},t_{1},t_{2}) = 2e^{2}\int\frac{\dd^{3}p}{(2\pi)^{3}} 
          & \left\lbrace1+\frac{px(px+\omega_{\vec{k}})+m^{2}}{p_{0}q_{0}}\right\rbrace
            n_{\text{F}}(q_{0})n_{\text{F}}(p_{0}) \nonumber \label{eq:2:pse_split_trans_anh} \\
   \times & \text{ }\ee^{-\ii(q_{0}+p_{0})(t_{1}-t_{2})} \ , \\
      \ii\Pi^{\text{PAC}}_{\text{T}}(\vec{k},t_{1},t_{2}) = 2e^{2}\int\frac{\dd^{3}p}{(2\pi)^{3}} 
          & \left\lbrace1+\frac{px(px+\omega_{\vec{k}})+m^{2}}{p_{0}q_{0}}\right\rbrace
            \left[1-n_{\text{F}}(q_{0})\right]\left[1-n_{\text{F}}(p_{0})\right] \nonumber \\
   \times & \text{ }\ee^{\ii(q_{0}+p_{0})(t_{1}-t_{2})} \ . \label{eq:2:pse_split_trans_pac}
  \end{align}
\end{subequations}
Here $p$ and $x$ denote the absolute value of the loop momentum, $\vec{p}$, and the cosine of the angle between $\vec{p}$ and $\vec{k}$, 
respectively, i.e., $\vec{p}\cdot\vec{k}=p\omega_{\vec{k}}x$. By making the substitutions $\vec{p}\rightarrow\vec{p}-\vec{k}$ and $x\rightarrow-x$ 
in (\ref{eq:2:pse_split_trans_abs}), it follows that this expression agrees with (\ref{eq:2:pse_split_trans_qbs}) for all values of $t_{1}$ and 
$t_{2}$. It is hence convenient to take these two contributions together as one single contribution describing (one-body) quark/antiquark 
Bremsstrahlung (BST), i.e., 
\begin{align}
 \label{eq:2:pse_split_trans_bst}
 \ii\Pi^{\text{BST}}_{\text{T}}(\vec{k},t_{1},t_{2}) = 4e^{2}\int\frac{\dd^{3}p}{(2\pi)^{3}} 
        & \left\lbrace1-\frac{px(px+\omega_{\vec{k}})+m^{2}}{p_{0}q_{0}}\right\rbrace
          n_{\text{F}}(q_{0})\left[1-n_{\text{F}}(p_{0})\right] \nonumber \\
 \times & \text{ }\ee^{-\ii(q_{0}-p_{0})(t_{1}-t_{2})} \ .
\end{align}
Accordingly, the photon number density (\ref{eq:2:photnum}) can be decomposed as
\begin{subequations}
 \label{eq:2:photnum_split}
 \begin{eqnarray}
  \left.2\omega_{\vec{k}}\frac{\dd^{6}n_{\gamma}(t)}{\dd^{3}x\dd^{3}k}\right|_{\text{BST}} 
    & = & \frac{1}{(2\pi)^{3}}\int_{-\infty}^{t}\dd t_{1}\int_{-\infty}^{t}\dd t_{2}\ii\Pi^{\text{BST}}_{\text{T}}(\vec{k},t_{1},t_{2})
          \ee^{\ii\omega_{\vec{k}}(t_{1}-t_{2})} \label{eq:2:photnum_split_bst} \ , \\
  \left.2\omega_{\vec{k}}\frac{\dd^{6}n_{\gamma}(t)}{\dd^{3}x\dd^{3}k}\right|_{\text{ANH}} 
    & = & \frac{1}{(2\pi)^{3}}\int_{-\infty}^{t}\dd t_{1}\int_{-\infty}^{t}\dd t_{2}\ii\Pi^{\text{ANH}}_{\text{T}}(\vec{k},t_{1},t_{2})
          \ee^{\ii\omega_{\vec{k}}(t_{1}-t_{2})} \label{eq:2:photnum_split_anh} \ , \\
  \left.2\omega_{\vec{k}}\frac{\dd^{6}n_{\gamma}(t)}{\dd^{3}x\dd^{3}k}\right|_{\text{PAC}} 
    & = & \frac{1}{(2\pi)^{3}}\int_{-\infty}^{t}\dd t_{1}\int_{-\infty}^{t}\dd t_{2}\ii\Pi^{\text{PAC}}_{\text{T}}(\vec{k},t_{1},t_{2})
          \ee^{\ii\omega_{\vec{k}}(t_{1}-t_{2})} \label{eq:2:photnum_split_pac} \ .
 \end{eqnarray}
\end{subequations}
That (\ref{eq:2:photnum_split_bst})-(\ref{eq:2:photnum_split_pac}) correspond to the contribution from the indicated processes 
can be seen by carrying out the multiplication of the respective expression for the photon self-energy with the factor $\ee^{\ii\omega_{\vec{k}}(t_{1}-t_{2})}$.
It follows from (\ref{eq:2:pse_split_trans_bst}) and (\ref{eq:2:pse_split_trans_qbs})-(\ref{eq:2:pse_split_trans_pac}) that this procedure gives rise to an oscillating 
behavior in $t_{1}-t_{2}$. The corresponding process can then be deduced from the specific oscillation frequency. Furthermore, when we show in 
appendix \ref{sec:appa} that each of the contributions (\ref{eq:2:photnum_split_bst})-(\ref{eq:2:photnum_split_pac}) can be written as the absolute 
square of a first-order QED transition amplitude, this interpretation also becomes evident from the underlying spinor structure.

In order to remove the artifacts encountered in \cite{Wang:2000pv,Wang:2001xh,Boyanovsky:2003qm}, we have to find an adequate ansatz for the 
fermion propagators (\ref{eq:2:propagators}). For this purpose, we take into account that the vacuum contribution to (\ref{eq:2:pse_loop}) 
occurs for all times, whereas the medium contributions only occur as long as the QGP is actually present. The former aspect is the reason why we have taken the 
initial time, i.e., the lower bound of the time integrals entering (\ref{eq:2:photnum}), to $-\infty$. The aforementioned time dependence is implemented into the 
fermion propagators (\ref{eq:2:propagators}) by introducing time dependent occupation numbers
\begin{equation}
 \label{eq:2:evol}
 n_{\text{F}}(E)\rightarrow n_{\text{F}}(E,t) = f(t)n_{\text{F}}(E) \ ,
\end{equation}
and replacing the fermion occupation numbers and the number of holes entering the fermion propagators (\ref{eq:2:propagators}) by their geometric 
mean from the different points of time, $t_{1}$ and $t_{2}$, i.e., 
\begin{subequations}
 \label{eq:2:evol_imp}
 \begin{eqnarray}
  n_{\text{F}}(E)   & \rightarrow & \sqrt{n_{\text{F}}(E,t_{1})n_{\text{F}}(E,t_{2})} \label{eq:2:evol_part} \ , \\
  1-n_{\text{F}}(E) & \rightarrow & \sqrt{\left[1-n_{\text{F}}(E,t_{1})\right]\left[1-n_{\text{F}}(E,t_{2})\right]} \label{eq:2:evol_hole} \ .
 \end{eqnarray}
\end{subequations}
By means of this procedure, the coincidence between (\ref{eq:2:pse_split_trans_qbs}) and (\ref{eq:2:pse_split_trans_abs}) is left unchanged. 
Moreover, the time evolution of the QGP is coupled to the interaction vertices. As we demonstrate in appendix \ref{sec:appa}, this ansatz ensures 
that (\ref{eq:2:photnum_split_bst})-(\ref{eq:2:photnum_split_pac}) can be written as an absolute square and, as a consequence, are positive 
(semi-)definite. Therefore, each of these contributions and thus the overall photon number density (\ref{eq:2:photnum}) cannot adopt unphysical 
negative values. Moreover, the absolute-square representation ensures that (\ref{eq:2:photnum_split_bst})-(\ref{eq:2:photnum_split_pac}) 
can be identified with the first-order QED process indicated in each case.

The crucial difference to \cite{Michler:2009hi} is that here we do not consider (\ref{eq:2:photnum}) at finite times, but in the limit 
$t\rightarrow\infty$ for free asymptotic states. In analogy to \cite{Michler:2012mg}, such states are obtained in this limit by 
introducing an adiabatic switching of the electromagnetic interaction according to the Gell-Mann and Low theorem, i.e., 
\begin{equation}
 \label{eq:2:adiabatic}
 \hat{H}_{\text{EM}} \rightarrow f_{\varepsilon}(t)\hat{H}_{\text{EM}} \ , \quad 
                                 \text{with} \quad f_{\varepsilon}(t)=\ee^{-\varepsilon|t|} \quad \text{and} \quad \varepsilon>0 \ .
\end{equation}
As a result, the time integrals entering (\ref{eq:2:photnum}) are effectively regulated by a factor of $\ee^{-\varepsilon|t_{i}|}$ with $i=1,2$. At 
the very end of our calculation, i.e., after taking the limit $t\rightarrow\infty$ in expression (\ref{eq:2:photnum}), we take the limit 
$\varepsilon\rightarrow 0$. As in \cite{Michler:2012mg}, the physical photon number density is thus defined as
\begin{equation}
 \label{eq:2:photnum_aspt}
 2\omega_{\vec{k}}\frac{\dd^{6}n_{\gamma}}{\dd^{3}x\dd^{3}k}
  = \lim_{\varepsilon\rightarrow0}\frac{1}{(2\pi)^{3}}\int_{-\infty}^{\infty}\dd t_{1}\int_{-\infty}^{\infty}\dd t_{2}f_{\varepsilon}(t_{1})f_{\varepsilon}(t_{2})
    \ii\Pi^{<}_{\text{T}}(\vec{k},t_{1},t_{2})\ee^{\ii\omega_{\vec{k}}(t_{1}-t_{2})} \ .
\end{equation}
We shall briefly demonstrate that (\ref{eq:2:photnum_aspt}) does not contain any unphysical contribution from the vacuum polarization. The 
latter is extracted from $\ii\Pi^{<}_{\text{T}}(\vec{k},t_{1},t_{2})$ by taking the limit $T\rightarrow0$ (which corresponds to the absence of 
the medium) and reads
\begin{equation}
 \label{eq:2:vacuum}
 \ii\Pi^{<}_{\text{\text{T},0}}(\vec{k},t_{1}-t_{2})
  = 2e^{2}\int\frac{\dd^{3}p}{(2\pi)^{3}}
     \left\lbrace
       1+\frac{px(px+\omega_{\vec{k}})+m^{2}}{p_{0}q_{0}}
     \right\rbrace
     \ee^{\ii(q_{0}+p_{0})(t_{1}-t_{2})} \ .
\end{equation}
Upon insertion of (\ref{eq:2:vacuum}) into (\ref{eq:2:photnum_aspt}), we obtain
\begin{eqnarray}
 \label{eq:2:vacuum_vanish}
 \left.\omega_{\vec{k}}\frac{\dd^{6}n_{\gamma}}{\dd^{3}x\dd^{3}k}\right|_{T\rightarrow0}
  & = &  \lim_{\varepsilon\rightarrow0}\frac{e^{2}}{(2\pi)^{3}}\int\frac{\dd^{3}p}{(2\pi)^{3}}
          \left\lbrace
           1+\frac{px(px+\omega_{\vec{k}})+m^{2}}{p_{0}q_{0}}
          \right\rbrace\cdot
          \left\lbrace
           \frac{2\varepsilon}{\varepsilon^{2}+\left(q_{0}+p_{0}+\omega_{\vec{k}}\right)^{2}}
          \right\rbrace^{2} \nonumber \\
  & \le & \lim_{\varepsilon\rightarrow0}\frac{4e^{2}}{(2\pi)^{3}}\int\frac{\dd^{3}p}{(2\pi)^{3}}
          \left\lbrace
           1+\frac{px(px+\omega_{\vec{k}})+m^{2}}{p_{0}q_{0}}
          \right\rbrace
           \cdot\frac{\varepsilon^{2}}{\left(q_{0}+p_{0}+\omega_{\vec{k}}\right)^{4}} \nonumber \\
  & =   & 0 \ ,
\end{eqnarray}
where we have taken into account that $q_{0}+p_{0}+\omega_{\vec{k}}>0$ in the second step.

\section{Numerical investigations and results}
\label{sec:numint}
\begin{figure}[htb]
 \begin{center}
  \includegraphics[height=5.0cm]{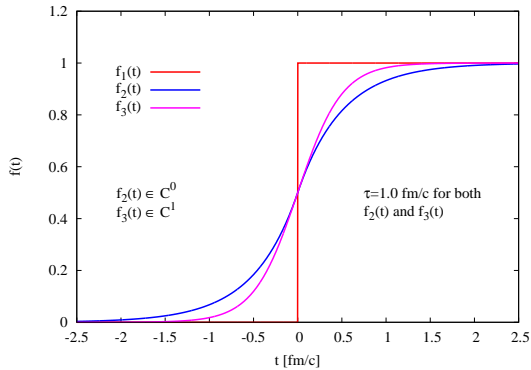}
  \caption{The time evolution of the QGP is modeled by different switching functions, $f_{i}(t)$.}
  \label{fig:switching_on}
 \end{center}
\end{figure}

In the previous section, we have presented the key features of our earlier model description on finite lifetime effects 
on the photon emission from a QGP. Now, we turn to our numerical investigations within this model approach. In this context we 
demonstrate that the consideration of the photon number density for free asymptotic states leads to UV integrable 
photon spectra if the time evolution of the quark/antiquark occupation numbers is modeled in a physically reasonable 
manner. For this purpose, we consider different switching functions, $f_{i}(t)$, for (\ref{eq:2:evol}). These switching functions 
are given by
\begin{subequations}
 \label{eq:3:switching_on}
 \begin{eqnarray}
  f_{1}(t) & = & \theta(t) \ , \\
  f_{2}(t) & = & \theta(t)-\frac{\text{sign}(t)}{2}\ee^{-2|t|/\tau} \ , \\
  f_{3}(t) & = & \frac{1}{2}\left[1+\tanh{\frac{2t}{\tau}}\right] \ ,
 \end{eqnarray}
\end{subequations}
and are depicted in Fig. \ref{fig:switching_on}.

$f_{1}(t)$ describes an instantaneous formation at $t=0$, whereas $f_{2}(t)$ and $f_{3}(t)$ describe a formation over a finite interval, $\tau$, 
in each case. Another difference between the latter two switching functions is that $f_{2}(t)$ is continuously differentiable once, whereas 
$f_{3}(t)$ is continuously differentiable infinitely many times. As in \cite{Michler:2009hi}, the photon self-energy, $\ii\Pi^{<}_{\text{T}}(\vec{k},t_{1},t_{2})$, 
is summed over the two light-quark flavors, up and down, such that $\sum_{f}e^{2}_{f}/e^{2}=5/9$, and the three colors. In order to avoid possible infrared 
and/or anticollinear singularities the quark/antiquark masses have been left finite, $m_{\text{u}}=m_{\text{d}}=0.01$ GeV.

\begin{figure}[htb]
 \begin{center}
  \includegraphics[height=5.0cm]{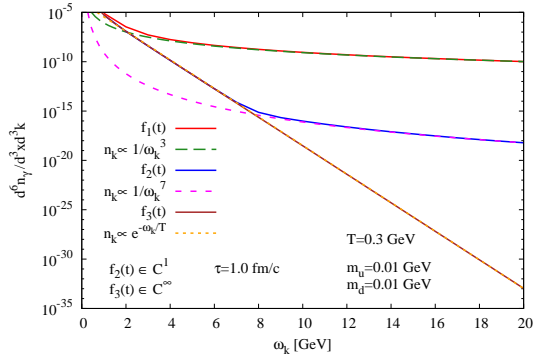}
  \caption{The scaling behavior of the photon number density in the UV domain is highly sensitive to the choice of $f(t)$. In particular, 
          it is rendered UV integrable if the QGP is assumed to be created over a finite interval of time, $\tau$.}
  \label{fig:photspec_comp_on}
 \end{center}
\end{figure}

Fig. \ref{fig:photspec_comp_on} compares the asymptotic photon spectra for the different switching functions, $f_{i}(t)$. For $f_{2}(t)$ and $f_{3}(t)$ 
a switching time of $\tau=1.0$ fm/c has been chosen.

For all three parameterizations, the loop integrals entering (\ref{eq:2:pse_split_trans_anh})-(\ref{eq:2:pse_split_trans_pac}) and (\ref{eq:2:pse_split_trans_bst})
are rendered finite by the Fermi-Dirac distribution function (\ref{eq:2:fermi-dirac}). In particular, this is also the case for (\ref{eq:2:pse_split_trans_pac}) 
since the contribution from the vacuum polarization characterized by the term proportional to $1$ is removed under the successive limits $t\rightarrow\infty$ and 
$\varepsilon\rightarrow0$, which also follows from Eqs. (\ref{eq:2:photnum_aspt})-(\ref{eq:2:vacuum_vanish}). For $f(t)=f_{1}(t)$ representing an instantaneous 
formation at $t=0$, the photon number density scales as $1/\omega^{3}_{\vec{k}}$ for large photon momenta, which means that the total number density and the total energy 
density of the emitted photons are logarithmically and linearly divergent, respectively. 

In contrast to \cite{Michler:2009hi}, however, this artifact is now fully removed if we turn form an instantaneous formation to a 
formation over a finite interval of time, $\tau$, representing a physically more reasonable scenario. For $f_{2}(t)$, which is continuously 
differentiable once, the photon number density is suppressed to $\propto 1/\omega^{7}_{\vec{k}}$, which means that the total photon number 
density and the total energy density are both UV finite. Moreover, if we turn from $f_{2}(t)$ to $f_{3}(t)$, which is continuously 
differentiable infinitely many times and hence represents the most physical scenario, the photon number density is suppressed even further 
to an exponential decay in $\omega_{\vec{k}}$. 

\begin{figure}[htb]
 \begin{center}
  \includegraphics[height=5.0cm]{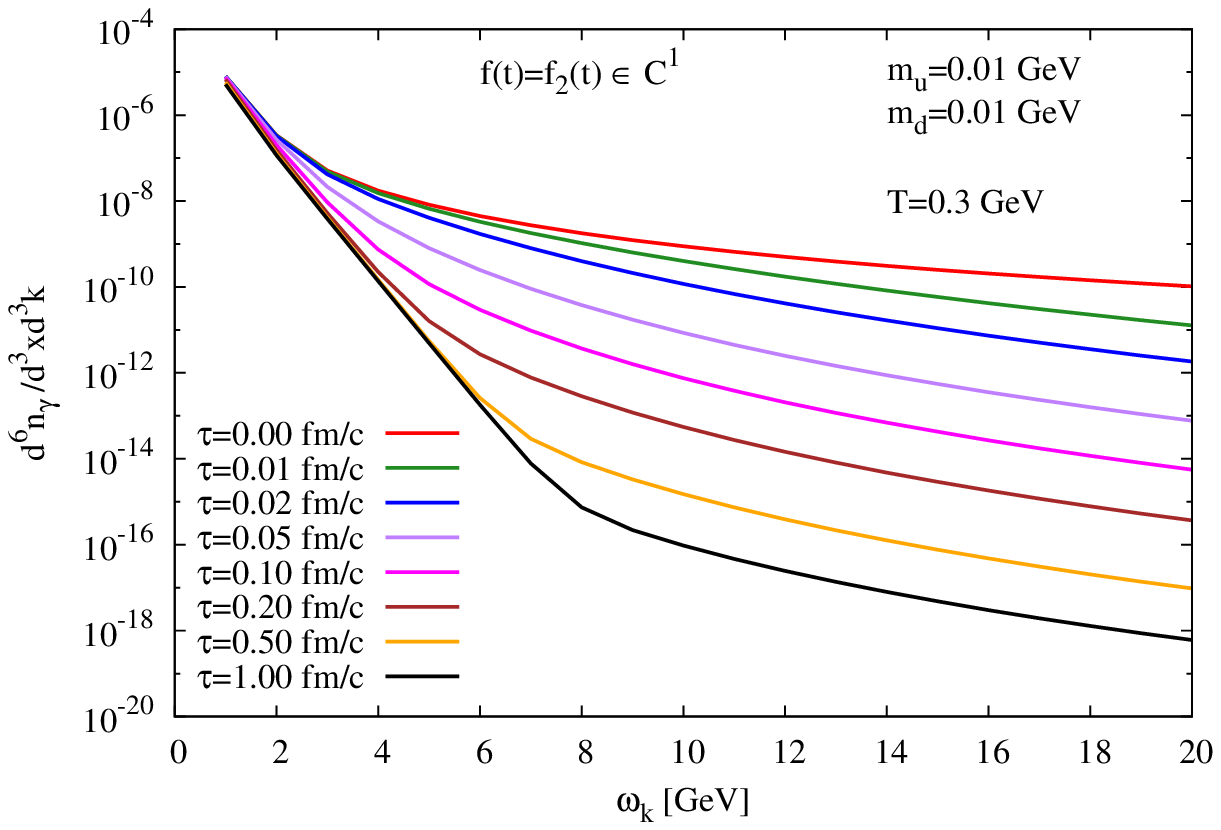}
  \includegraphics[height=5.0cm]{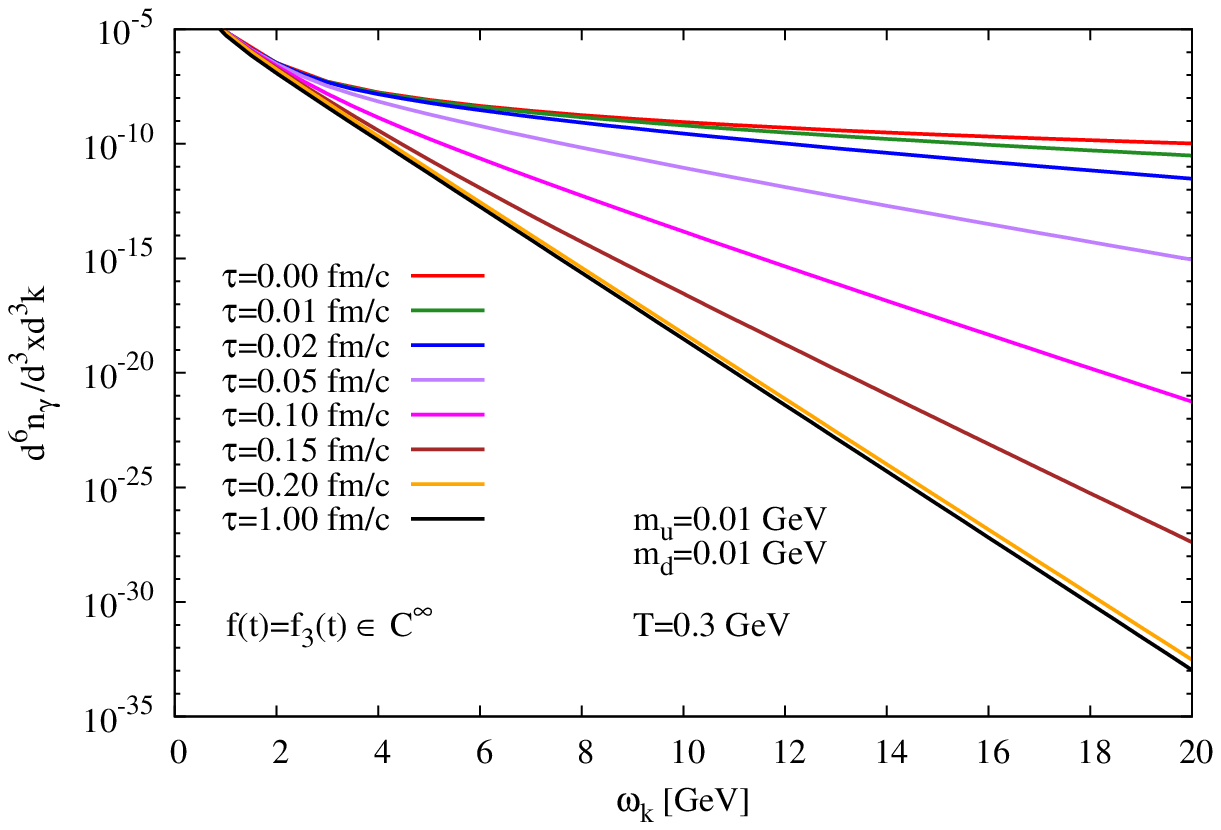}
  \caption{For both $f_{2}(t)$ (left panel) and $f_{3}(t)$ (right panel) the photon spectrum for $f_{1}(t)$ is reproduced in the limit 
           $\tau\rightarrow0$. For $f_{2}(t)$, the suppression of the photon number density with respect to the instantaneous case is the stronger 
           the larger $\tau$ is chosen. To the contrary, this quantity seems to converge against some finite value with increasing $\tau$ for 
           $f_{3}(t)$ with the slope of the photon spectrum then given by $\beta$.}
  \label{fig:photspec_tauzero_on}
 \end{center}
\end{figure}

One remarkable feature in this context is that the slope of the photon spectrum, i.e., the energy scale over which the photon number density decreases by 
a factor of $1/\ee$ for large $\omega_{\vec{k}}$, coincides with $\beta=1/T$ for $\tau=1.0$ fm/c. This suggests that the photon spectrum starts looking thermal with $\tau$ increasing 
from $0$ (where it coincides with the one for $f_{1}(t)$) if the quark/antiquark occupation numbers are switched on according to $f_{3}(t)$. A comparison of the photon 
spectra for different switching times, which is provided in Fig. \ref{fig:photspec_tauzero_on}, supports this.

Nevertheless, in this context the exact dependence of the photon number density on the switching time, $\tau$, is counterintuitive for 
$f_{3}(t)$: If the quark/antiquark occupation numbers are switched on according to $f_{2}(t)$, the suppression of the photon number density 
in the UV domain with respect to the instantaneous case is the stronger the larger $\tau$ is chosen, i.e., the more slowly the formation of the QGP 
is assumed to take place. Furthermore, $f_{2}(t)$ reproduces the photon spectrum for the instantaneous case in the limit $\tau\rightarrow0$, 
as it must be. The latter is also the case if the quark/antiquark occupation numbers are switched on by means of $f_{3}(t)$. In the limit 
$\tau\rightarrow\infty$, however, the photon number density seems to converge against some finite value and, as a consequence, to become 
independent of $\tau$. To the contrary, one would expect intuitively that in this limit said quantity disappears. Then one effectively 
has a static plasma such that first-order QED processes become kinematically impossible.

\begin{figure}[htb]
 \begin{center}
  \includegraphics[height=5.0cm]{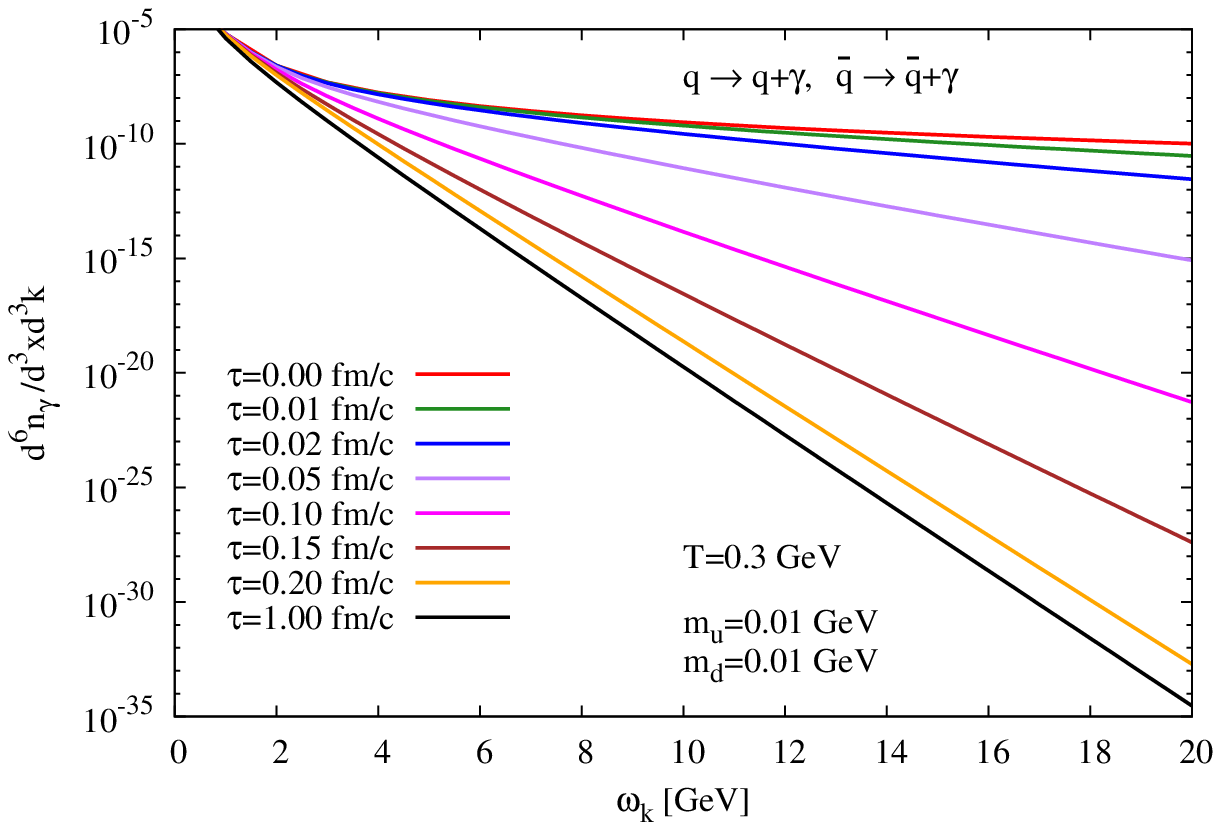}
  \includegraphics[height=5.0cm]{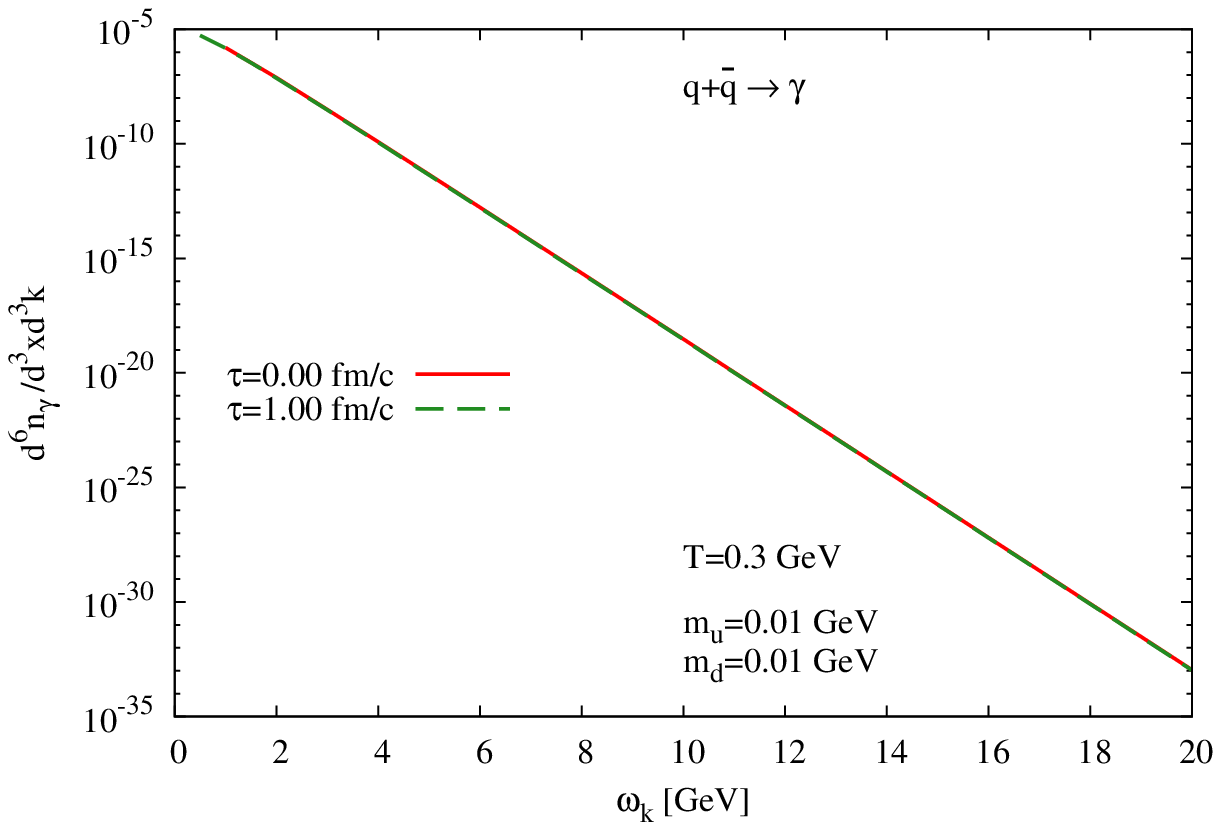}
  \caption{Dependence of the contributions from quark/antiquark Bremsstrahlung (left panel) and from quark-antiquark pair 
           annihilation into a single photon (right panel) on the switching time, $\tau$, for $f_{3}(t)$. The contribution from 
           quark/antiquark Bremsstrahlung seems to saturate in the limit $\tau\rightarrow\infty$ with the slope of the spectrum then 
           given by $\beta$. Furthermore, the photon spectrum arising from quark-antiquark pair annihilation into a single photon 
           seems to be entirely independent of $\tau$ and exhibits the same slope.}
  \label{fig:photspec_tauzero_split_01}
 \end{center}
\end{figure}

In the following, we demonstrate that the latter is indeed the case. For this purpose, we first consider the photon spectra for 
each of the processes contributing to (\ref{eq:2:photnum_aspt}) separately. Fig. \ref{fig:photspec_tauzero_split_01} shows the photon 
spectra arising from quark/antiquark Bremsstrahlung and quark-antiquark pair annihilation into a single photon.

\begin{figure}[htb]
 \begin{center}
  \includegraphics[height=5.0cm]{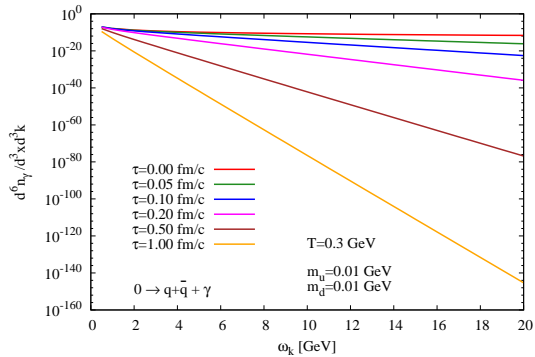}
  \caption{For the contribution arising from the spontaneous creation of a quark-antiquark pair together with a photon out of the vacuum, it is 
           evident that its suppression with respect to the instantaneous case is the stronger the more slowly ($\tau$ increasing) the formation 
           of the QGP is assumed to take place and that it eventually disappears in the limit $\tau\rightarrow\infty$.}
  \label{fig:photspec_tauzero_split_02}
 \end{center}
\end{figure}

We see that the inverse slope of the photon spectrum arising from quark/antiquark Bremsstrahlung seems to converge against $\beta$ with increasing $\tau$, 
and that the photon number density appears to converge to a finite value in the limit $\tau\rightarrow\infty$ for a given photon energy, $\omega_{\vec{k}}$. 
Furthermore, the photon spectrum arising from quark-antiquark pair annihilation into a single photon seems to be independent of $\tau$ with its 
slope also given by $\beta$. To the contrary, for the contribution from the spontaneous creation of a quark-antiquark pair together 
with a photon out of the vacuum one can infer from Fig. \ref{fig:photspec_tauzero_split_02} that its suppression with respect to the 
instantaneous case is the stronger the larger $\tau$ is chosen and that it accordingly disappears in the limit $\tau\rightarrow\infty$.

\begin{figure}[htb]
 \begin{center}
  \includegraphics[height=5.0cm]{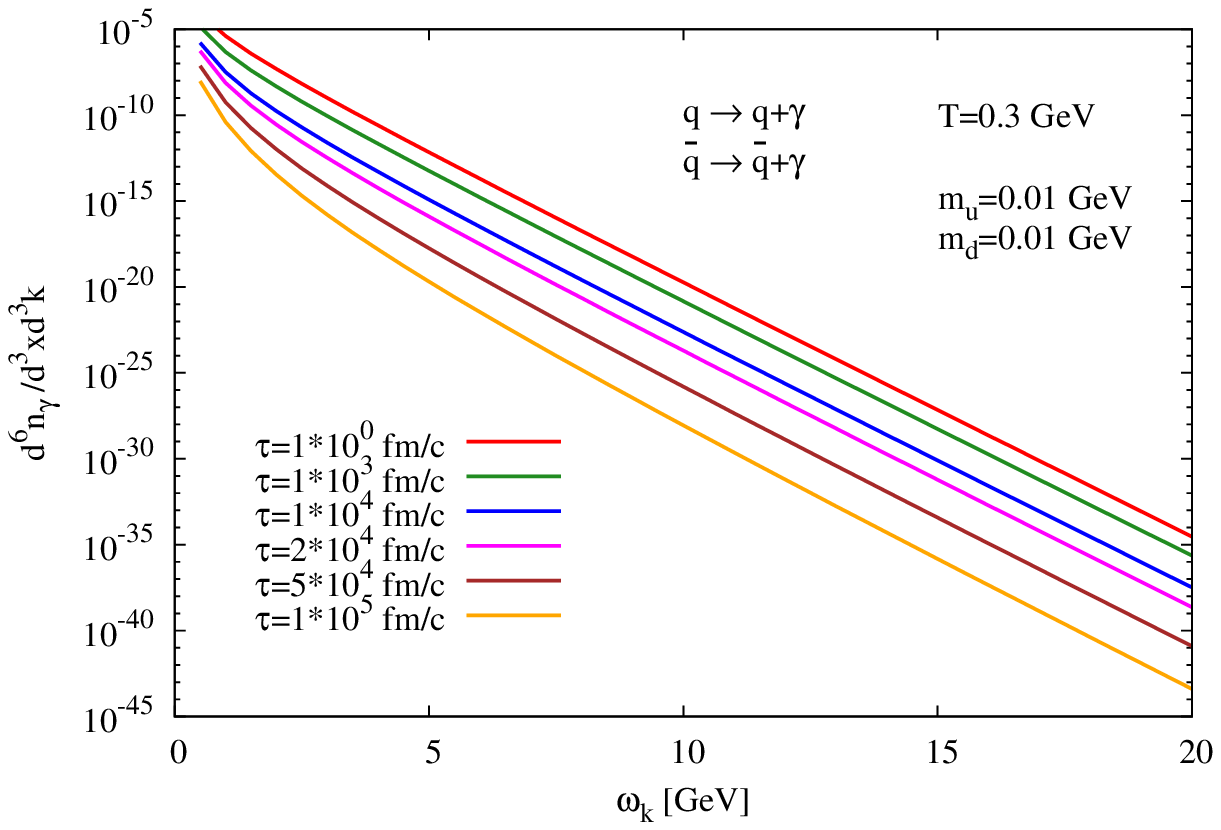}
  \includegraphics[height=5.0cm]{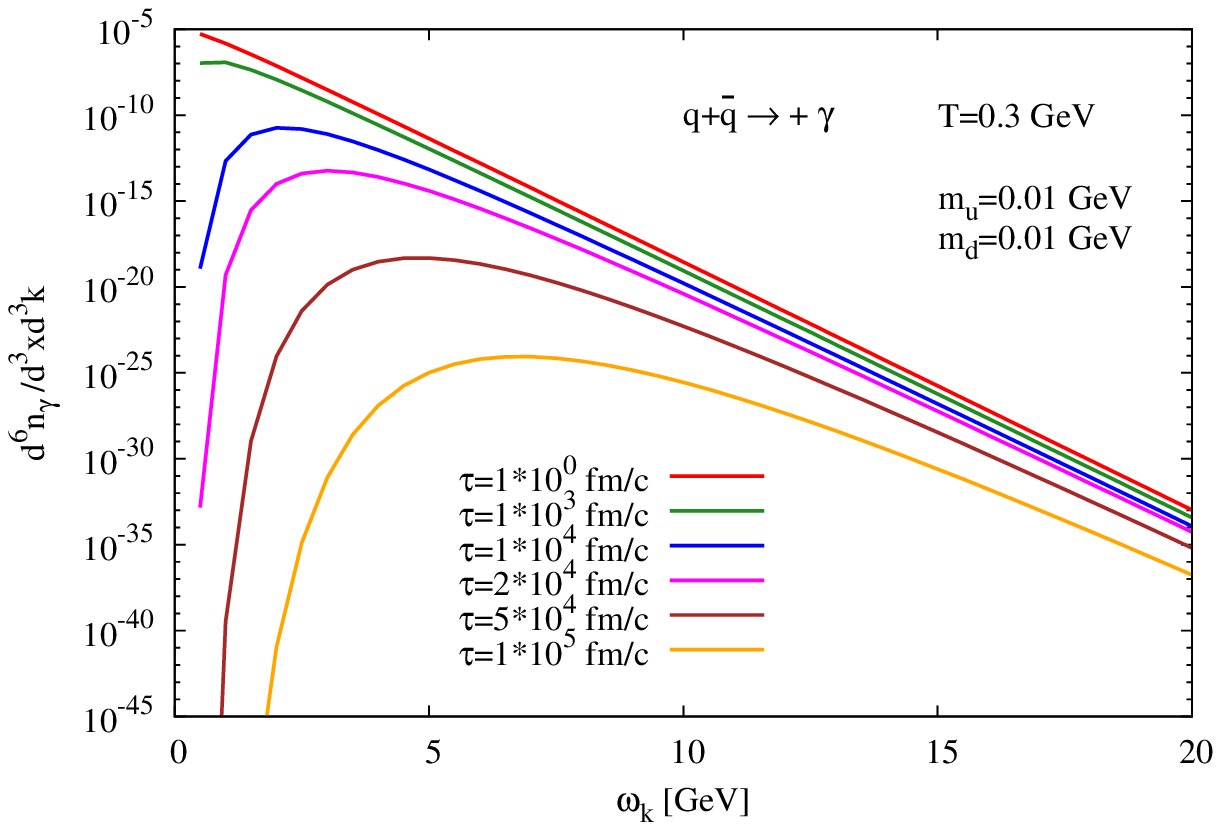}
  \caption{If the photon spectra arising from quark/antiquark Bremsstrahlung (left panel) and quark-antiquark pair annihilation into a single photon are
           considered for switching times exceeding the expected formation time of the QGP by several orders of magnitude, one sees that both contributions 
           also vanish in the limit $\tau\rightarrow\infty$.}
  \label{fig:photspec_largetau}
 \end{center}
\end{figure}

This implies that the apparent saturation of the overall photon number density in the limit $\tau\rightarrow\infty$ results from the 
contributions from quark/antiquark Bremsstrahlung and quark-antiquark pair annihilation into a single photon. As one expects intuitively, 
however, these contributions (and hence the overall photon number density) do not saturate but also vanish in the above limit. In order to 
see this, one has to consider them for switching times that exceed the expected (from the phenomenological point of view) formation time of the QGP  
of $\tau_{\text{QGP}}\simeq1.0$ fm/c \cite{Heinz:2001xi} by several orders of magnitude. This can be inferred from Fig. \ref{fig:photspec_largetau}. For the 
contribution from the spontaneous creation of a quark-antiquark pair together with a photon out of the vacuum, to the contrary, the expected 
disappearance in the limit $\tau\rightarrow\infty$ already becomes visible for switching times being of the same order of magnitude as the 
expected formation time of the QGP.

We shall give an explanation for how such a different dependence on $\tau$ comes about for the individual contributions to 
(\ref{eq:2:photnum_aspt}). For this purpose, we take into account that each of them is given by a loop integral over the different 
loop-momentum modes contributing to the respective underlying first-order QED process. Each of these modes is characterized by a specific 
formation time. For the individual first-order QED processes, these formation times read
\begin{subequations}
 \label{eq:3:formation_times}
 \begin{eqnarray}
  \tau_{\text{BST}}(\vec{p},\vec{k}) & = & \frac{2\pi}{\left|q_{0}-p_{0}-\omega_{\vec{k}}\right|} \ , \label{eq:3:formation_times_bst} \\
  \tau_{\text{ANH}}(\vec{p},\vec{k}) & = & \frac{2\pi}{q_{0}+p_{0}-\omega_{\vec{k}}} \ , \label{eq:3:formation_times_anh} \\
  \tau_{\text{PAC}}(\vec{p},\vec{k}) & = & \frac{2\pi}{q_{0}+p_{0}+\omega_{\vec{k}}} \ , \label{eq:3:formation_times_pac}
 \end{eqnarray}
\end{subequations}
with the denominators denoting the required virtuality, i.e. the `offshellness', of the respectively considered process. In equation (\ref{eq:3:formation_times_bst}) 
we have taken into account that the frequency $q_{0}-p_{0}-\omega_{\vec{k}}$ is negative definite. 

For a specific photon-emission mode that contributes to a particular process to be suppressed with respect to the instantaneous case, the switching time, $\tau$, has to be chosen 
significantly larger than the formation time of the considered mode. The reason is that then the QGP appears to be static for this mode by which the associated process  
becomes effectively kinematically impossible. When considering the contribution to the photon number density from this particular process, this implies that $\tau$ has to be 
chosen significantly larger than the formation times of all contributing emission modes such that the disappearance of respective contribution in the limit $\tau\rightarrow\infty$ 
becomes evident.

On the other hand, for the contributions from quark/antiquark Bremsstrahlung and  quark-antiquark pair annihilation into a single photon the formation times 
of the collinear ($x=1$) and the anticollinear modes ($x=-1$) in the domain $p\le\omega_{\vec{k}}$, respectively, exhibit formation times exceeding the expected 
formation time of the QGP by several orders of magnitude, which can be read from Table \ref{tab:formation}. As a consequence, the switching time has to be 
chosen significantly larger than these formation times and hence by several orders of magnitude larger than the expected formation time of the QGP such that 
it becomes visible that the contributions from quark/antiquark Bremsstrahlung and quark-antiquark pair annihilation into a single photon vanish in 
the limit $\tau\rightarrow\infty$. For the sake of clarity, we would like to stress again that $x$ denotes the cosine of the angle between the photon momentum, $\vec{k}$, 
and the fermion-loop momentum, $\vec{p}$, i.e., $\vec{p}\cdot\vec{k}=p\omega_{\vec{k}}x$, such that the collinear and the anticollinear photon-emission modes are characterized 
by $x=1$ and $x=-1$, respectively.
\begin{table}
 \begin{center}
  \begin{tabular}{c||c|c c c||c|l l l||}
                               &         & \multicolumn{3}{c||}{$\tau_{\text{BST}}(\vec{p},\vec{k})$ [fm/c]} &         & \multicolumn{3}{c||}{$\tau_{\text{ANH}}(\vec{p},\vec{k})$ [fm/c]}    \\
                               & p [GeV] & x=1.0              & & x=0.9                                        & p [GeV] & x=-1.0              & & x=-0.9                                     \\
    \hline
    $\omega_{\vec{k}}=5.0$ GeV &  2.0    & $3.52\cdot 10^{5}$ & & $8.70\cdot 10^{0}$                           & 1.0     & $2.01\cdot 10^{4}$  & & $1.02\cdot 10^{1}$                             \\
                               &  4.0    & $9.05\cdot 10^{5}$ & & $5.58\cdot 10^{0}$                           & 2.0     & $3.02\cdot 10^{4}$  & & $3.97\cdot 10^{0}$                             \\
                               &  6.0    & $1.66\cdot 10^{6}$ & & $4.55\cdot 10^{0}$                           & 3.0     & $3.02\cdot 10^{4}$  & & $1.95\cdot 10^{0}$                             \\
                               &  8.0    & $2.61\cdot 10^{6}$ & & $4.04\cdot 10^{0}$                           & 4.0     & $2.01\cdot 10^{4}$  & & $1.02\cdot 10^{0}$                             \\
                               & 10.0    & $3.77\cdot 10^{6}$ & & $3.73\cdot 10^{0}$                           & 5.0     & $1.26\cdot 10^{0}$  & & $5.62\cdot 10^{-1}$                            \\
                               & 12.0    & $5.13\cdot 10^{6}$ & & $3.52\cdot 10^{0}$                           & 6.0     & $6.28\cdot 10^{-1}$ & & $3.45\cdot 10^{-1}$  
  \end{tabular}
  \caption{Formation times of the collinear modes for the process of quark/antiquark Bremsstrahlung (left part) and of the anticollinear modes for 
           the process of quark-antiquark pair annihilation into a single photon (right part) for $\omega_{\vec{k}}=5.0$ GeV and $m_{\text{u}}=m_{\text{d}}=0.01$ GeV. 
           One can see that the formation times of the collinear modes and of the anticollinear modes in the domain $p\le\omega_{\vec{k}}$ exceed 
           the expected formation time of the QGP by several orders of magnitude. In this context, it is particularly remarkable that the formation times in turn 
           decrease by several orders of magnitude if they are considered for modes outside these domains, i.e., if one decreases $x$ from $1.0$ to $0.9$ for the 
           contribution from quark/antiquark Bremsstrahlung or if one either increases $p$ from some $p\le\omega_{\vec{k}}$ to $6.0$ GeV or $x$ from $-1.0$ to $-0.9$ 
           for the contribution from quark-antiquark pair annihilation into a single photon. In each case, the formation time is of the same or even in a smaller 
           order of magnitude than the expected formation time of the QGP.}
  \label{tab:formation}
 \end{center}
\end{table}

\begin{figure}[htb]
 \begin{center}
  \includegraphics[height=5.0cm]{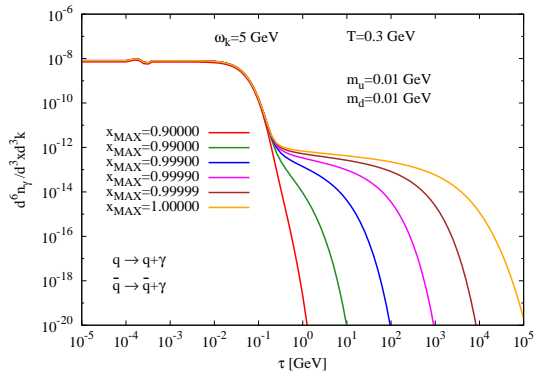}
  \caption{Dependence of the contribution arising from quark/antiquark Bremsstrahlung on the switching time, $\tau$, for different upper bounds, 
           $x_{\text{MAX}}$, for the integration over $\dd x$. If the collinear modes are excluded, this contribution decreases much faster 
           with increasing $\tau$. As it must be, the actual decreasing behavior is reproduced if $x_{\text{MAX}}$ is increased back to $1$.}
  \label{fig:irange_cut_bst}
 \end{center}
\end{figure}

This can be seen by restricting the integration range over $\dd^{3}p$ such that the collinear (quark/antiquark Bremsstrahlung) and the 
anti-collinear modes for $p\le\omega_{\vec{k}}$ (quark-antiquark pair annihilation into a single photon) are excluded. In this case, 
the respective contribution decreases much faster with increasing $\tau$ and, depending on the exact restriction of the integration range, it becomes 
visible that both of them disappear for large $\tau$ already for values around $1$ fm/c. For the contribution from quark/antiquark Bremsstrahlung, this 
can be seen in Fig. \ref{fig:irange_cut_bst}, where the upper bound of the integration over $\dd x$ is varied. If we choose $x_{\text{MAX}}=0.9$ 
such that the collinear modes are excluded, the contribution from quark/antiquark Bremsstrahlung decreases much faster with increasing $\tau$. In particular, 
it becomes evident that it disappears in the limit $\tau\rightarrow\infty$ even if $\tau$ is of the order of $1$ fm/c, which coincides with the expected formation 
time of the QGP. On the other hand, if $x_{\text{MAX}}$ is increased gradually back to $1$ the collinear modes are successively re-included such that the 
decrease of the Bremsstrahlung contribution with increasing $\tau$ is delayed accordingly.

\begin{figure}[htb]
 \begin{center}
  \includegraphics[height=5.0cm]{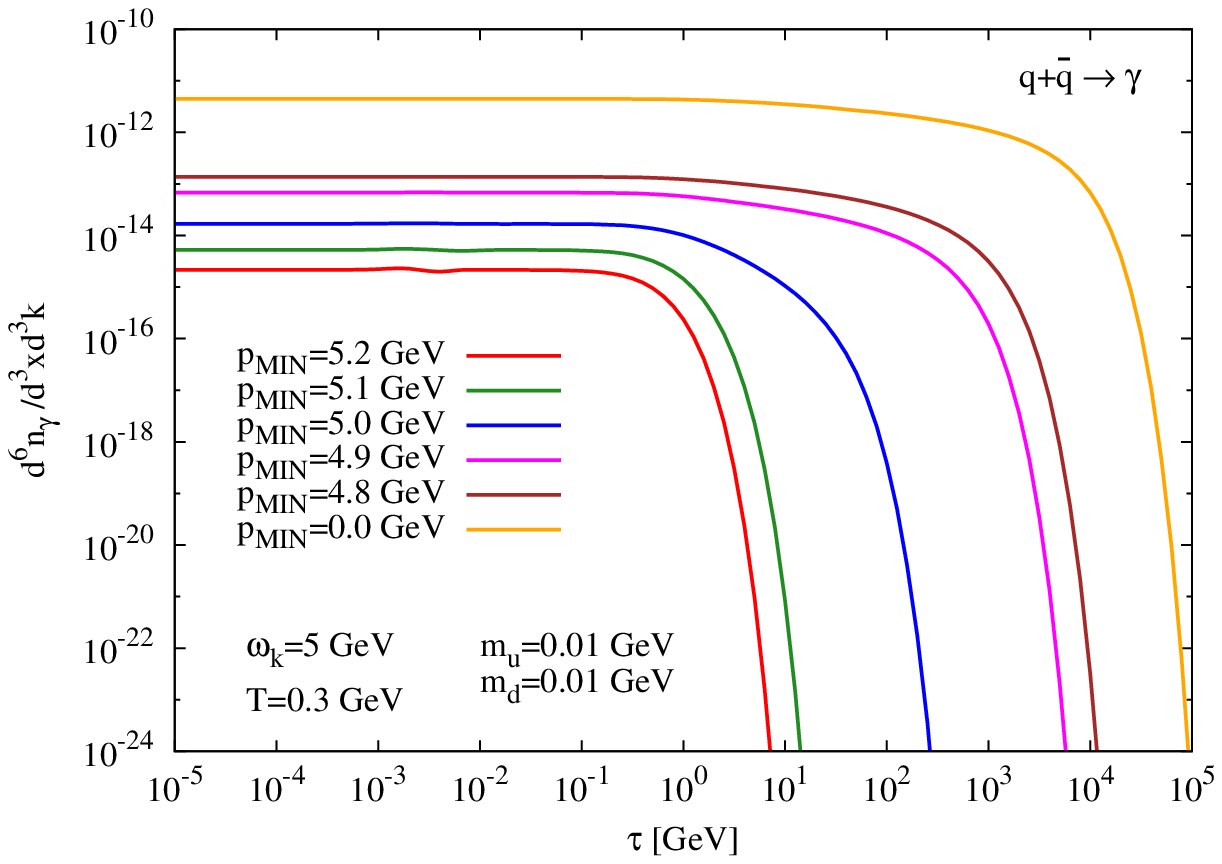}
  \includegraphics[height=5.0cm]{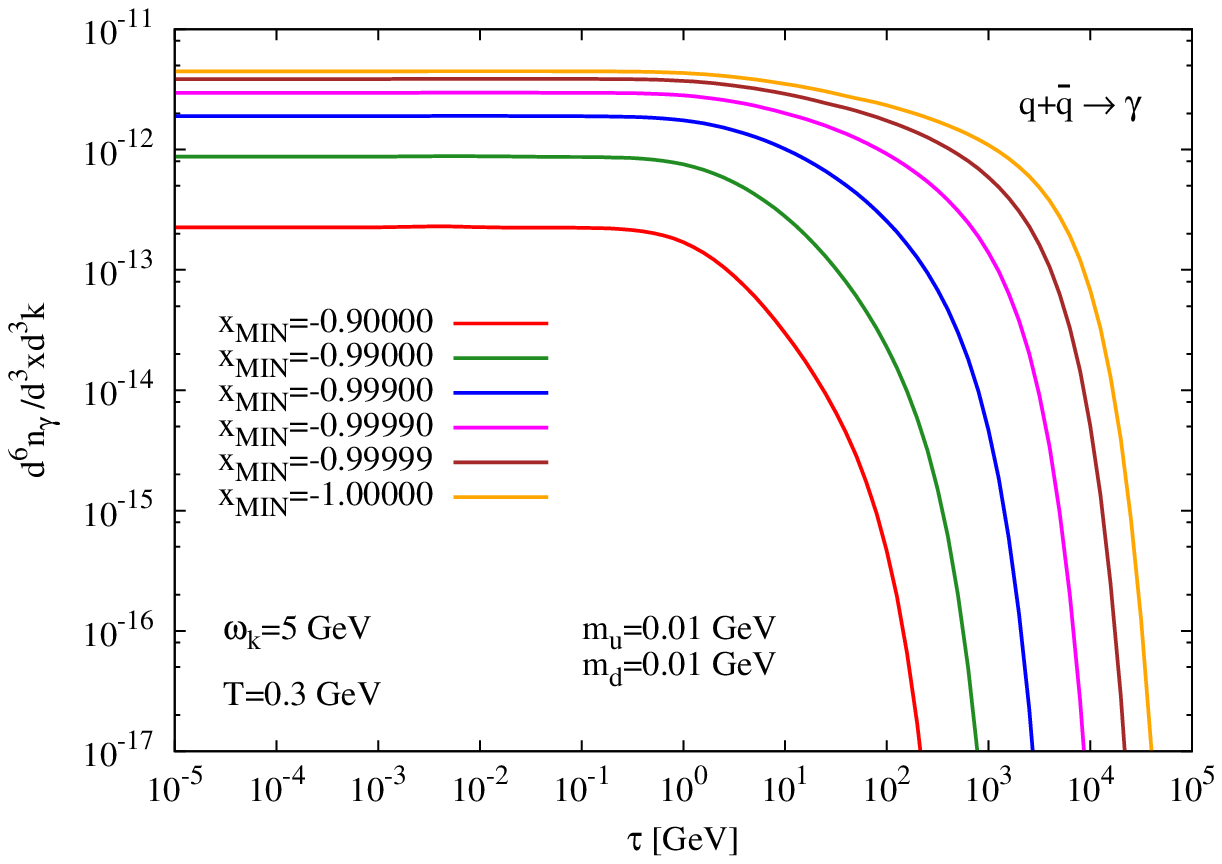}
  \caption{Dependence of the contribution arising from quark-antiquark pair annihilation into a single photon on the switching time, $\tau$, with 
           different lower bounds, $p_{\text{MIN}}$, for integration over $\dd p$ (left panel) and different lower bounds, $x_{\text{MIN}}$, for 
           the integration over $\dd x$ (right panel). If the anticollinear modes at $p\le\omega_{\vec{k}}$ are excluded this contribution also 
           decreases much faster with increasing $\tau$. As expected, the actual decreasing behavior is reproduced if we decrease $p_{\text{MIN}}$    
           back to $1.0$ GeV and $x_{\text{MIN}}$ back to $1.0$, respectively.}
  \label{fig:irange_cut_anh}
 \end{center}
\end{figure}

Analogously, the contribution from quark-antiquark pair annihilation into a single photon decreases considerably faster with increasing $\tau$ if either the anticollinear 
modes or the modes for which $p\le\omega_{\vec{k}}$ are excluded. This is shown in Fig. \ref{fig:irange_cut_anh}, where the lower bound of the integrations 
over $\dd p$ and $\dd x$ are varied from $5.2$ GeV down to $0$ GeV (left panel) and from $-0.9$ down to $-1.0$ (right panel), respectively. If we choose either 
$p_{\text{MIN}}=5.2$ GeV or $x_{\text{MIN}}=-0.9$ the anticollinear modes in the domain $p\le\omega_{\vec{k}}$ are excluded, and the contribution from quark-antiquark pair annihilation 
into a single photon decreases much faster with increasing $\tau$ than it does for a full integration over $\dd^{3}p$. As a consequence, it becomes evident that this contribution disappears 
in the limit $\tau\rightarrow\infty$ already if $\tau$ is chosen around $10$ fm/c. If $p_{\text{MIN}}$ and $x_{\text{MIN}}$ are gradually decreased back to 
$0.0$ GeV and $-1.0$, respectively, the anticollinear modes from the range $p\le\omega_{\vec{k}}$ are re-included and the decrease of the of the pair-annihilation 
contribution is effectively delayed.

This shows that the apparent saturation of the contributions from quark/antiquark Bremsstrahlung and quark-antiquark pair annihilation into a 
single photon for $\tau$ being varied from $0-1$ fm/c results from the large formation times of the collinear and 
anticollinear modes in the range $p\le\omega_{\vec{k}}$, respectively. To the contrary, for the spontaneous creation of a quark-antiquark 
pair together with a photon out of the vacuum the formation times of all contributing modes are bounded by 
\begin{equation}
 \label{eq:3:form_est}
 \tau_{\text{PAC}}(\vec{p},\vec{k}) \le \frac{2\pi}{2m_{\text{u,d}}+\omega_{\vec{k}}} \ ,
\end{equation}
for a specific photon energy, $\omega_{\vec{k}}$, such that the contribution from this process decreases much faster with increasing $\tau$. 
Accordingly, its vanishing in the limit $\tau\rightarrow\infty$ manifests itself already for switching times of the same order of magnitude 
as the formation time of the QGP.

We have seen that the contributions from quark/antiquark Bremsstrahlung and quark-antiquark pair annihilation into a single photon
decrease much faster with increasing $\tau$ if the collinear and the anticollinear modes at $p\le\omega_{\vec{k}}$ are excluded 
from the integration over $\dd^{3}p$ in each case. Accordingly, said modes lead to an enhancement of the respective contribution to the overall 
photon number density by several orders of magnitude for the physically motivated choice of $\tau\simeq\tau_{\text{QGP}}\simeq1.0$ fm/c. Such an enhancement, which eventually 
might turn into a (anti-) collinear divergence $m_{\text{u,d}}\rightarrow0$, requires an HTL-resummation of the quark/antiquark propagators. 
This effectively assigns the quarks and antiquarks a thermal mass. The full quark/antiquark mass hence reads
\begin{equation}
 \label{eq:3:full_mass}
 m^{\text{full},2}_{\text{u,d}} = m^{\text{bare},2}_{\text{u,d}}+m^{2}(T) \ ,
\end{equation}
where we have chosen $m^{\text{bare}}_{\text{u,d}}=0.01$ GeV, and the thermal component, $m(T)$, given by
\begin{equation}
 \label{eq:3:thermal_mass}
 m^{2}(T)=\frac{4\pi\alpha_{s}}{3}\left(N_{\text{c}}+\frac{N_{\text{f}}}{2}\right)T^{2} \ .
\end{equation}
Here $N_{\text{c}}$ and $N_{\text{f}}$ denote the number of colors and flavors, respectively. If we consider three colors and the 
two light-quark flavors, up and down, expression (\ref{eq:3:thermal_mass}) turns into
\begin{equation}
 m^{2}(T)=\frac{16\pi\alpha_{s}}{3}T^{2} \ .
\end{equation}
For a temperature of $T=0.3$ GeV and $\alpha_{s}\approx0.3$, the thermal component of the quark/antiquark mass is of the order of several hundred 
MeV and hence significantly larger than the bare component. This in turn implies that if the thermal component of (\ref{eq:3:full_mass}) is taken into 
account the actual formation times of the collinear modes and the anticollinear modes at $p\le\omega_{\vec{k}}$ contributing to the processes quark/antiquark 
Bremsstrahlung and quark/antiquark pair annihilation into a single photon, respectively, are significantly smaller compared to the case in which only the 
bare component is considered. This is shown in Table \ref{tab:masses}.
\begin{table}
 \begin{center}
  \begin{tabular}{c||c|c c c||c|l l l||}
                               &         & \multicolumn{3}{c||}{$\tau_{\text{BST}}(\vec{p},\vec{k})$ [fm/c]} &         & \multicolumn{3}{c||}{$\tau_{\text{ANH}}(\vec{p},\vec{k})$ [fm/c]}    \\
                               & p [GeV] & $m_{\text{u,d}}=m^{\text{bare}}_{\text{u,d}}$ & & $m_{\text{u,d}}=m^{\text{full}}_{\text{u,d}}$ & p [GeV] & $m_{\text{u,d}}=m^{\text{bare}}_{\text{u,d}}$ & & $m_{\text{u,d}}=m^{\text{full}}_{\text{u,d}}$ \\
    \hline
    $\omega_{\vec{k}}=5.0$ GeV &  2.0    & $3.52\cdot 10^{5}$ & & $1.61\cdot 10^{1}$        & 1.0     & $2.01\cdot 10^{4}$  & & $6.81\cdot 10^{0}$                               \\
                               &  4.0    & $9.05\cdot 10^{5}$ & & $4.05\cdot 10^{1}$        & 2.0     & $3.02\cdot 10^{4}$  & & $4.81\cdot 10^{0}$                               \\
                               &  6.0    & $1.66\cdot 10^{6}$ & & $7.38\cdot 10^{1}$        & 3.0     & $3.02\cdot 10^{4}$  & & $5.60\cdot 10^{-1}$                              \\
                               &  8.0    & $2.61\cdot 10^{6}$ & & $1.16\cdot 10^{2}$        & 4.0     & $2.01\cdot 10^{4}$  & & $2.06\cdot 10^{-1}$                              \\
                               & 10.0    & $3.77\cdot 10^{6}$ & & $1.67\cdot 10^{2}$        & 5.0     & $1.26\cdot 10^{0}$  & & $1.25\cdot 10^{-1}$                              \\
                               & 12.0    & $5.13\cdot 10^{6}$ & & $2.27\cdot 10^{2}$        & 6.0     & $6.28\cdot 10^{-1}$ & & $8.94\cdot 10^{-2}$    
  \end{tabular}
  \caption{If the thermal component of the quark/antiquark mass is taken into account, the formation times of the collinear modes and the 
           anticollinear modes at $p\le\omega_{\vec{k}}$ contributing to the processes of quark/antiquark Bremsstrahlung and quark-antiquark pair 
           annihilation into a single photon, respectively, are significantly smaller compared to the case where only the bare component is 
           considered. For the thermal component of the quark/antiquark mass, we have chosen $T=0.3$ GeV and $\alpha_{s}\approx0.3$, which
           implies that $m(T)\approx 0.67$ GeV.}
  \label{tab:masses}
 \end{center}
\end{table}

\begin{figure}[htb]
 \begin{center}
  \includegraphics[height=5.0cm]{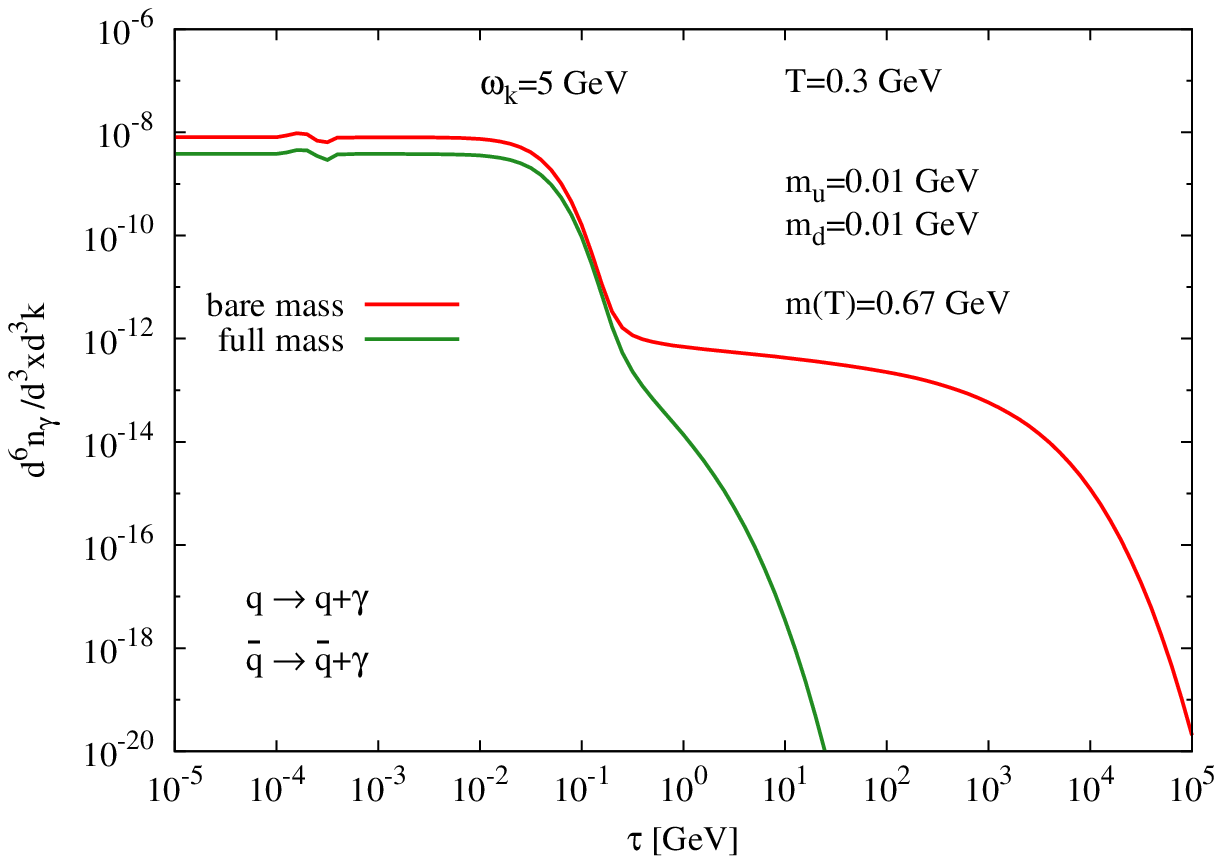}
  \includegraphics[height=5.0cm]{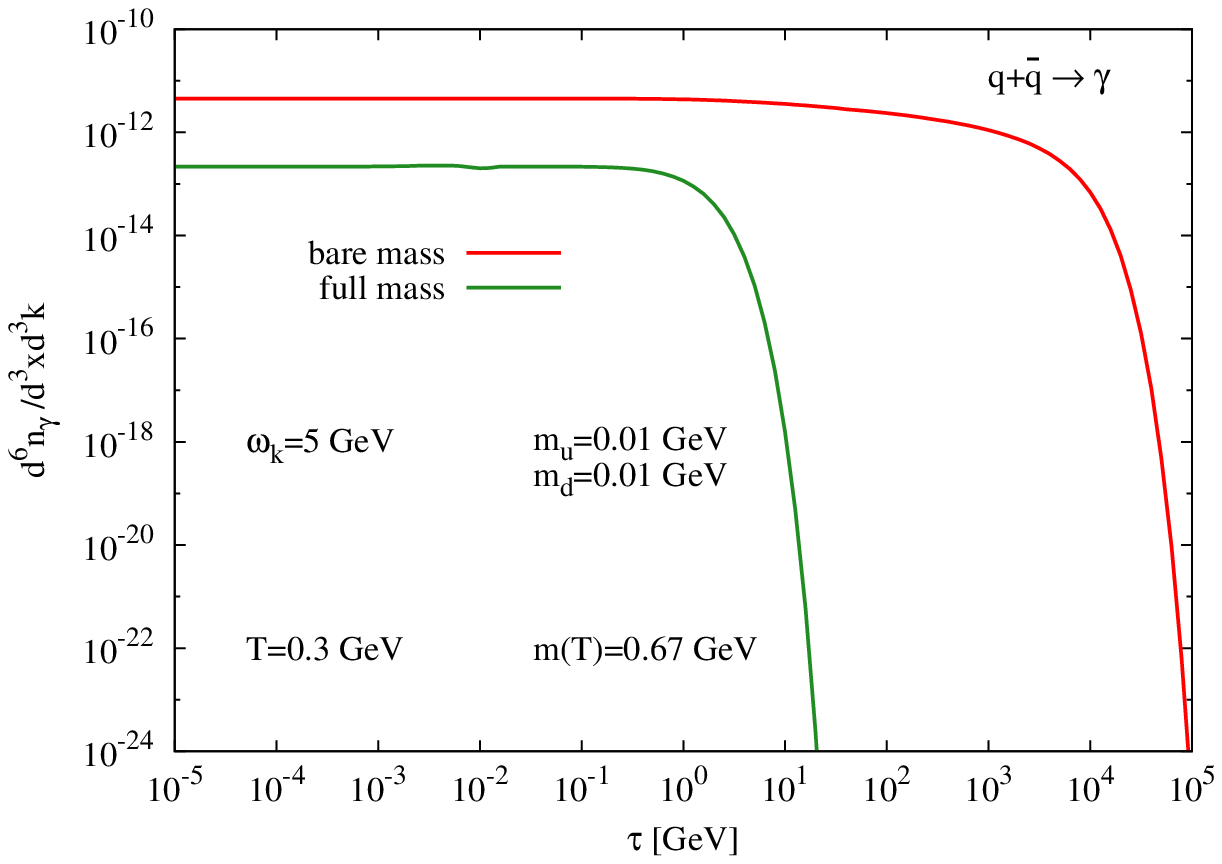}
  \caption{Dependence of the contributions from quark/antiquark Bremsstrahlung (left panel) and from quark-antiquark pair annihilation 
           into a single photon (right panel) on the switching time, $\tau$, for the bare and the full quark/antiquark masses. If the 
           thermal component of the mass is included, both contributions decrease much faster with increasing $\tau$ compared to the 
           case where only the bare masses are considered.}
  \label{fig:massdep}
 \end{center}
\end{figure}

One hence expects that the contributions from quark/antiquark Bremsstrahlung and quark/antiquark pair annihilation into a single photon then 
accordingly decrease considerably faster with increasing $\tau$. As a consequence, the disappearance of these contributions for $\tau\rightarrow\infty$ 
should become evident even if $\tau$ is chosen from the same order of magnitude as the expected formation time of the QGP. One can infer from 
Fig. \ref{fig:massdep} that this is indeed the case.

In this work, we have only presented results on the scenario in which the quark/antiquark occupation numbers are switched on 
and maintained, but not the scenario in which they are switched back off after a certain period of time, $\tau_{\text{L}}$, to take into account 
the finite lifetime of the QGP during a heavy-ion collision. The reason is that for the latter scenario the principle sensitivity of the photon number 
density on the switching function, $f(t)$, and the switching time, $\tau$, is as in the one presented here. Firstly, the photon number density 
again scales as $1/\omega^{3}_{\vec{k}}$ in the UV domain if the quark/antiquark occupation numbers are switched on and off instantaneously, with this 
artifact being removed if both switchings take place over a finite interval of time, $\tau$, instead. In particular, the photon spectrum again starts 
looking thermal with increasing $\tau$ if we consider a switching function being continuously differentiable infinitely many times. For such a 
switching function, the photon number density again seems to converge against some finite value for $\tau\rightarrow\infty$, where this apparent 
saturation can again be traced back to the large formation times of the collinear modes and the anticollinear modes at $p\le\omega_{\vec{k}}$ for 
the processes of quark/antiquark Bremsstrahlung and quark-antiquark pair annihilation into a single photon, respectively.

When comparing the exact dependence of the asymptotic photon spectra within our (revisited) model description to those from \cite{Michler:2012mg}, 
the fact that the slope of the overall photon spectrum converges against the inverse temperature, $\beta$, (provided that the quark/antiquark occupation numbers are 
switched on according to $f_{3}(t)$) might seem unphysical at first. In \cite{Michler:2012mg}, we have investigated the photon emission arising 
from a change of the quark/antiquark mass. We have seen that the slope of the resulting photon spectrum increases with the transition time of the 
quark/antiquark mass if the time evolution of the latter is modeled by a function being continuously differentiable infinitely many times. Hence, 
one might expect a similar dependence of the photon spectra on the switching time, $\tau$, within our revisited model description if the 
quark/antiquark occupation numbers are switched according to $f_{3}(t)$ since this function is also continuously differentiable infinitely many 
times.

Here it is important to point out, however, that within our (revisited) model description, we always switch on the same distribution function for 
the quarks and antiquarks for all switching functions, $f(t)$, and, in particular, for all considered switching times, $\tau$. To the contrary, in 
\cite{Michler:2012mg} we pursue a first-principle approach in which the quark/antiquark occupation numbers are determined by solving the Dirac 
equation with a time dependent mass. This has the direct consequence that the quark/antiquark occupation numbers decrease exponentially with 
increasing momentum, $p$, and that the slope of the respective spectrum increases with the transition time (provided that the mass function is 
continuously differentiable infinitely many times). This in turn manifests itself in form of a very similar sensitivity of the asymptotic photon 
spectrum on this time. To the contrary, such a specific dependence does not occur within our model description since by construction the latter 
features quark/antiquark occupation numbers which solely depend on the temperature and are hence independent of $\tau$.

\section{Comparison to leading-order thermal photon production}
\label{sec:thermcomp}
The investigations from \cite{Wang:2000pv,Wang:2001xh,Boyanovsky:2003qm} indicated that non-equilibrium photon production arising from 
first-order QED processes possibly dominates over leading-order thermal photon emission in the UV domain. On the other hand, these 
investigations came along with the mentioned artifacts, which in turn questions the explanatory power of the comparison performed therein. 
Since the artifacts from \cite{Wang:2000pv,Wang:2001xh,Boyanovsky:2003qm} have been resolved to a satisfactory extend within this work, 
we again perform a comparison to leading-order thermal photon production in order to get a more significant picture. Here we note again that 
the contributions from first-order QED processes to photon production vanish in a static thermal equilibrium such that there the first non-trivial 
contribution starts at two-loop order. Since a loop expansion does not coincide with a coupling-constant expansion, resummations of so-called 
ladder diagrams are necessary in order to obtain the thermal rate at second order in the perturbative coupling constants, i.e., at linear 
order in $\alpha_{e}$ and at linear order in $\alpha_{s}$ \cite{Arnold:2001ms}.

Within the scope of our investigations on chiral photon production \cite{Michler:2012mg}, we have already made a rather rudimentary 
comparison to leading-order thermal photon emission by simply integrating the rate from \cite{Arnold:2001ms} over the assumed lifetime of 
the chirally restored phase at constant temperature. This comparison indicated that (first-order) non-equilibrium photon production is 
subdominant compared to leading-order thermal production for photon energies $\omega_{\vec{k}}\gtrsim1.0$ GeV.

Since the actual question from \cite{Wang:2000pv,Wang:2001xh,Boyanovsky:2003qm} on the role of finite lifetime effects on direct photon 
emission from a QGP is readdressed within this work, we now perform a more detailed comparison. In this context we take into account that the QGP 
as it occurs in a heavy-ion collision is not a static medium but instead expands and cools down over a finite interval of time before it 
hadronizes finally. To begin with, the time dependence of the temperature effectively leads to a time dependent photon-production 
rate, i.e., 
\begin{equation}
 \label{eq:4:rate_thermal}
 \frac{\dd^{7}n_{\gamma}}{\dd^{4}x\dd^{3}k} = \frac{\dd^{7}n_{\gamma}(T(t))}{\dd^{4}x\dd^{3}k} \equiv \frac{\dd^{7}n_{\gamma}(t)}{\dd^{4}x\dd^{3}k} \ .
\end{equation}
In order to obtain the overall photon number accessible to experiment, one has to convolute (\ref{eq:4:rate_thermal}) with the 
time dependent volume, $V_{\text{QGP}}(t)$, of the expanding QGP from the initial time, $t_{0}$, at which the QGP has thermalized until 
the time $t_{\text{had}}$, at which the full hadronic phase is reached. This leads to
\begin{equation}
 \label{eq:4:photnum_eq}
 \left.\frac{\dd^{3}n_{\gamma}}{\dd^{3}k}\right|_{\text{eq.}} = \int_{t_{0}}^{t_{\text{had}}}\dd t V_{\text{QGP}}(t)\frac{\dd^{7}n_{\gamma}(t)}{\dd^{4}x\dd^{3}k} \ .
\end{equation}
For the time evolution of the volume and the temperature of the QGP, we consider the same fireball model that has been used in \cite{vanHees:2011vb} for $0-20$ \% 
central Au+Au collisions at $200$ AGeV.

When calculating the overall photon number arising from the first-order non-equilibrium contributions, we multiply our asymptotic 
photon number density directly with the initial volume of the QGP, i.e.,
\begin{equation}
 \label{eq:4:photnum_noneq}
 \left.\frac{\dd^{3}n_{\gamma}}{\dd^{3}k}\right|_{\text{non-eq.}} = V_{\text{QGP}}(t_{0})\frac{\dd^{6}n_{\gamma}}{\dd^{3}x\dd^{3}k} \ .
\end{equation}
The reason is that the first-order non-equilibrium photon production occurs during the formation of the QGP, which we model by the switching-on 
of the quark/antiquark occupation numbers. The adiabatic switching-off of the electromagnetic interaction then removes the artificial 
contributions occurring at finite times. As a consequence, the asymptotic photon number density has also to be computed for the initial 
temperature, $T_{0}$.

For our numerical analysis, we chose the same values for the parameters of the fireball model as done in \cite{vanHees:2011vb}. In particular, 
we assume an initial volume of the QGP of $V_{\text{QGP}}(t_{0})=73.76$ $\text{fm}^{3}$ which from the underlying equation of state leads to an 
initial temperature of $T_{0}=0.36$ GeV. For the photon numbers emerging from first-order non-equilibrium emission process, we consider a  
switching time of $\tau=1.0$ fm/c. The latter are considered both for the bare and the full quark/antiquark mass. As a consequence, the thermal mass is taken 
with respect to the initial temperature, which accordingly to (\ref{eq:3:thermal_mass}) leads to $m(T_{0})=0.81$ GeV.

\begin{figure}[htb]
 \begin{center}
  \includegraphics[height=5.0cm]{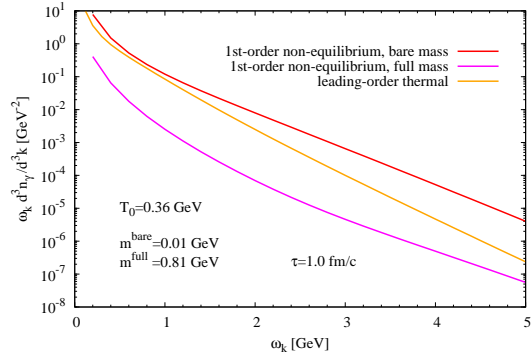}
  \caption{Comparison of first-order non-equilibrium photon production to leading-order thermal production. If one takes into account the full  
           quark/antiquark mass, the former is subdominant for $\omega_{\vec{k}}=1-5$ GeV.}
  \label{fig:thermcomp}
 \end{center}
\end{figure}

Fig.~\ref{fig:thermcomp} compares the photon spectra for first-order non-equilibrium production to those for leading-order thermal production. If we only 
take into account the bare component of the quark/antiquark mass, the non-equilibrium photon emission exceeds the thermal emission by one one order 
of magnitude for photon energies $\omega_{\vec{k}}\gtrsim1.0$ GeV. At first sight this seems to support the qualitative picture from \cite{Wang:2000pv,Wang:2001xh,Boyanovsky:2003qm}. 
Here it is important to point out, however, that the photon spectrum for the full quark/antiquark mass is the more realistic one since the 
included thermal component effectively provides the required HTL-resummation of the (anti-)collinear photon-emission modes. In this case we see 
that in contrast to \cite{Wang:2000pv,Wang:2001xh,Boyanovsky:2003qm}, the photon numbers arising from first-order non-equilibrium processes are clearly below 
those arising from leading-order thermal photon production for $\omega_{\vec{k}}=1-5$ GeV.

On the other hand, the photon spectrum emerging from leading-order thermal contributions features a steeper decay that the one from the first-order 
non-equilibrium contributions since it incorporates the entire time evolution of the temperature of the QGP and not only its initial temperature, $T_{0}$. 
This in turn indicates that non-equilibrium photon production becomes dominant somewhere above $\omega_{\vec{k}}=5$ GeV.

This does, however, effectively not change the principle idea that direct photon production from the QGP phase during a heavy-ion collision can be 
addressed by integrating the leading-order thermal rate on a hydrodynamic background (quasi-static calculation) and that a full dynamic treatment 
is not crucial quantitatively. Comprehensive comparisons of the contributions from the different sources of direct photon emission to the overall photon spectra 
measured in RHIC and LHC experiments \cite{Turbide:2003si,Turbide:2005fk,Gale:2009gc} have shown that medium contributions from the hadronic phase 
dominate in the infrared (IR) domain, whereas the photon emission arising from initial nucleon-nucleon scatterings and jet-medium interactions 
outshine the medium contributions both from the QGP and the hadronic phase in the UV domain. To the contrary, a dominance of the medium contribution 
from the QGP phase could only possibly be observed at intermediate photon energies with the exact range increasing with the collision energy. On the other hand, 
our investigations have shown that for these intermediate energies, leading-order thermal photon production clearly dominates over the  
first-order non-equilibrium one.

The principal reason why the contributions form initial nucleon-nucleon scatterings and jet-medium interactions dominate over the pure medium contributions 
from the QGP and the subsequent hadronic phase in the UV domain is that the photon spectra from the former two sources flatten into a power-law decay, 
whereas those from the latter feature an exponential decay. Such an exponential decay is also observed for the photon spectra arising from first-order 
non-equilibrium production from the QGP. This implies that even though this photon production starts to dominate over the leading-order thermal one at photon 
energies $\omega_{\vec{k}}\gtrsim5$ GeV such that a quasi-static description strictly speaking becomes invalid in this domain, this does not effectively matter 
since the medium contributions from the QGP are outshone by the contributions from initial nucleon-nucleon scatterings and jet-medium interactions in any case.

\section{Remarks on the importance of free asymptotic states}
\label{sec:aspt}
We would like to stress again that the exact sequence of limits, i.e., taking \textit{first} $t\rightarrow\infty$ and 
\textit{then} $\varepsilon\rightarrow0$, is crucial to eliminate a possible unphysical contribution from the vacuum polarization and, 
in general, to obtain a UV integrable photon number density from the medium contributions to $\ii\Pi^{<}_{\text{T}}(\vec{k},t_{1},t_{2})$. 
If one interchanges both limits, i.e., if one first takes $\varepsilon\rightarrow0$ at some finite time, $t$, it can be shown 
\cite{Michler:2012mg} that the contribution from the vacuum polarization does not vanish, but instead turns into
\begin{equation}
 \label{eq:5:limexch_vac}
 \left.\omega_{\vec{k}}\frac{\dd^{6}n_{\gamma}}{\dd^{3}x\dd^{3}k}\right|_{T\rightarrow0}
  = \frac{e^{2}}{(2\pi)^{3}}\int\frac{\dd^{3}p}{(2\pi)^{3}}
          \left\lbrace
           1+\frac{px(px+\omega_{\vec{k}})+m^{2}}{p_{0}q_{0}}
          \right\rbrace
     \frac{1}{\left(q_{0}+p_{0}+\omega_{\vec{k}}\right)^{2}} \ .
\end{equation}
Since the integration measure, $\dd^{3}p$, contributes an additional factor of $p^{2}$ to the integrand, the loop integral is linearly 
divergent for a given photon energy, $\omega_{\vec{k}}$. Furthermore, since (\ref{eq:5:limexch_vac}) is time independent, it persists under the 
subsequent limit $t\rightarrow\infty$.

\begin{figure}[htb]
 \begin{center}
  \includegraphics[height=5.0cm]{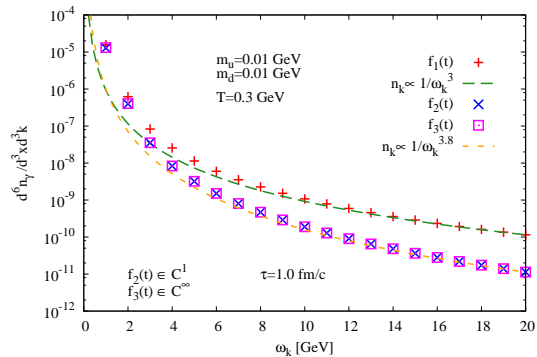}
  \caption{If both limits are interchanged, the UV scaling behavior of the photon number density solely changes from 
           $\propto1/\omega^{3}_{\vec{k}}$ to $\propto1/\omega^{3.8}_{\vec{k}}$ when turning from an instantaneous 
           switching to a switching over a finite interval of time, $\tau$.}
  \label{fig:photspec_limexch}
 \end{center}
\end{figure}

On the other hand, since (\ref{eq:5:limexch_vac}) is time independent and hence already present before any medium contributions to (\ref{eq:2:photnum}) 
can appear, one might still argue that it can be identified with the virtual cloud of the vacuum and accordingly needs to be subtracted since it is unobservable. 
The reason for this time independence, which suggests such an identification, is that in contrast to \cite{Wang:2000pv,Wang:2001xh,Boyanovsky:2003qm}, our description takes into 
account that the vacuum contribution to the photon self-energy always occurs, whereas for the medium contributions this is only the case as long as the QGP is actually present. 
After subtracting the divergent vacuum contribution and taking the subsequent limit $t\rightarrow\infty$, however, one in general 
still encounters the problem that the photon number density arising from the remaining medium contributions to the photon self-energy is not integrable in the UV domain. 
This is shown in Fig. \ref{fig:photspec_limexch}.

\begin{figure}[htb]
 \begin{center}
  \includegraphics[height=5.0cm]{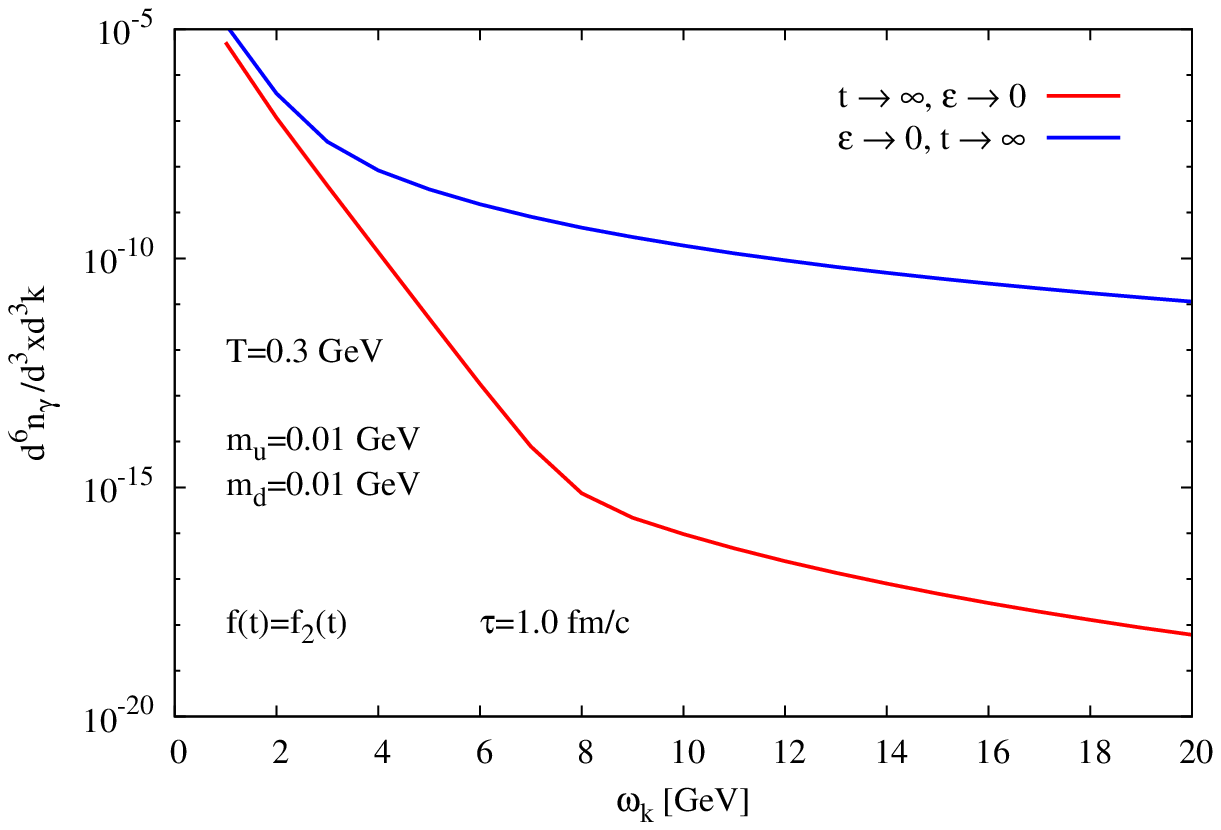}
  \includegraphics[height=5.0cm]{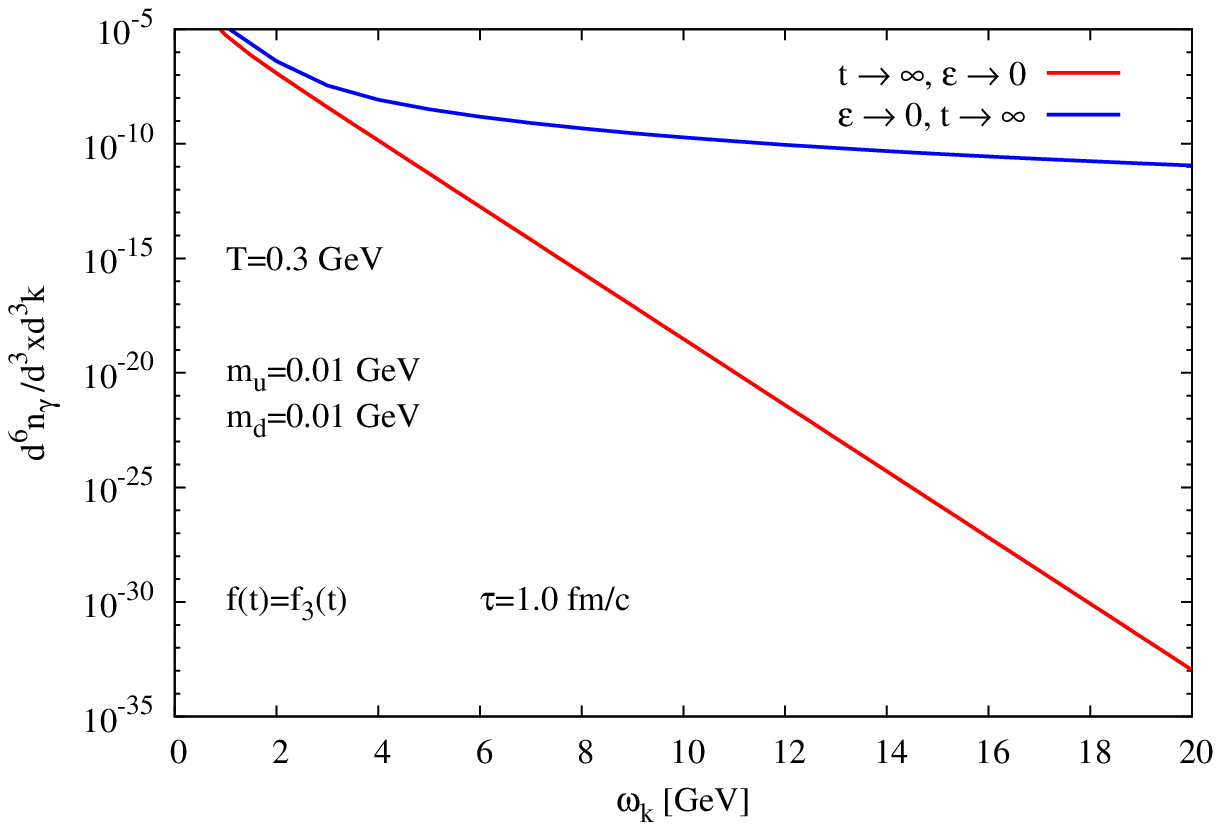}
  \caption{Comparison of the photon number density for the correct and the interchanged sequence of limits for $f(t)=f_{2}(t)$ (left panel) 
           and $f(t)=f_{3}(t)$ (right panel). For the interchanged sequence of limits the photon number density exceeds the value for the 
           correct sequence of limits by several orders of magnitude in the UV domain in both cases.}
  \label{fig:photspec_seqcomp}
 \end{center}
\end{figure}

As for the correct sequence of limits the photon number density scales $\propto 1/\omega^{3}_{\vec{k}}$ for $f_{1}(t)$. If we turn from $f_{1}(t)$ 
to $f_{2}(t)$ or $f_{3}(t)$, we see, however, that the photon number density is suppressed to only a slightly steeper decay 
$\propto 1/\omega^{3.8}_{\vec{k}}$. For that reason, only the total number density of the radiated photons is UV finite, whereas their total 
energy density remains UV divergent. In particular, the thus obtained photon number density exceeds the value for the correct sequence of 
limits by several orders of magnitude, which is displayed in Fig. \ref{fig:photspec_seqcomp}.

Here it is important to point out once more that the interpretation of (\ref{eq:2:photnum}) as a photon number density is only justified 
in the limit $t\rightarrow\pm\infty$ for finite $\varepsilon$ since only then the electromagnetic field is asymptotically non-interacting. At finite 
times, higher order (Fock) contributions to the photon number density still persist due to the remote interactions. This implies that taking first 
$\varepsilon\rightarrow0$ at some finite time, $t$, is not correct, as then we would have an \emph{interacting} electromagnetic field such that the 
interpretation of (\ref{eq:2:photnum}) as a photon number density is not justified. Moreover, such an interpretation remains doubtful even in the limit 
$t\rightarrow\infty$. Since we would have taken $\varepsilon\rightarrow0$ before the electromagnetic field would not evolve into a non-interacting one. 
The same conceptual problem occurs when only using an adiabatic switching-on of the electromagnetic interaction for $t\rightarrow-\infty$ but no adiabatic 
switching-off for $t\rightarrow\infty$. Such a procedure has been suggested in \cite{Serreau:2003wr} to describe initial correlations at some $t=t_{0}$ evolving from an 
uncorrelated initial state at $t\rightarrow-\infty$.

\section{Summary, conclusions and outlook}
\label{sec:conclusions}
In this work, we have investigated the role of finite lifetime effects on the photon emission from a rapidly created quark-gluon plasma (QGP) during a heavy-ion 
collision. We have essentially revisited our earlier model description \cite{Michler:2009hi}, in which we simulate the time evolution 
of the QGP by time dependent quark/antiquark occupation numbers in the photon self-energy. In contrast to \cite{Michler:2009hi}, we have not 
considered the photon number density at finite times, but for free asymptotic states, as the former is ill-defined. In analogy to \cite{Michler:2012mg}, we have seen that 
this procedure does eliminate a possible unphysical contribution from the vacuum polarization and, moreover, leads to a UV integrable photon number 
density. This result confirms the conjecture that the artifacts encountered in \cite{Wang:2000pv,Wang:2001xh,Boyanovsky:2003qm,Michler:2009hi} arise from an 
inconsistent definition of the photon number density at finite times. Consequently, our investigations again support the corresponding concern raised in 
\cite{Fraga:2003sn,Fraga:2004ur} towards \cite{Wang:2000pv,Wang:2001xh,Boyanovsky:2003qm}.

When switching the quark/antiquark occupation numbers by an analytic function, which represents the physically most reasonable scenario, we have seen that 
the photon number density apparently converges to a finite value for large $\tau$ if the latter is chosen of the same order of magnitude as the 
(phenomenologically) expected formation time of the QGP, which amounts to $\tau_{\text{QGP}}\simeq 1.0$ fm/c. In order to see that the photon number density actually 
vanishes in the limit $\tau\rightarrow\infty$, the switching time has to be chosen larger than $\tau_{\text{QGP}}$ by several orders of magnitude.

We have shown that this apparent saturation results from the contributions describing quark/antiquark Bremsstrahlung and quark-antiquark 
pair annihilation into a single photon. In contrast to the spontaneous creation of a quark-antiquark pair together with a photon out of the 
vacuum, both of these processes feature contributions from individual photon-emission modes for which the formation times exceed $\tau_{\text{QGP}}$ 
by several orders of magnitude. In particular, these modes are the collinear ones for the process of quarks/antiquark Bremsstrahlung and the 
anticollinear ones at $p\le\omega_{\vec{k}}$ (with $p$ denoting the absolute value of the loop momentum) for the process of quark-antiquark pair annihilation into 
a single photon. On the other hand, the switching time, $\tau$, has to be chosen significantly larger than the formation
time of \emph{all} modes contributing to a specific process such that the disappearance of the contribution from this process (and hence of 
the overall photon number density) in the limit $\tau\rightarrow\infty$ becomes evident. This can be seen by excluding said modes from 
the integration range over $\dd^{3}p$. In this case, the contributions from quark/antiquark Bremsstrahlung and quark-antiquark pair annihilation into 
a single photon decrease much faster with increasing $\tau$. In particular, it becomes evident that these contributions vanish for $\tau\rightarrow\infty$ 
even if $\tau$ is chosen to be of the same order of magnitude as $\tau_{\text{QGP}}$.

On the other hand, said (anti-)collinear photon-emission modes lead to a significant (by several orders of magnitude) enhancement of the respective contributions 
to the overall photon number density for $\tau\simeq\tau_{\text{QGP}}$. Strictly speaking, such an enhancement requires 
an HTL resummation of the quark/antiquark propagators, by which the quarks and antiquarks are effectively assigned a thermal mass. This thermal 
component of the quark/antiquark mass is by $1-2$ orders of magnitude larger than the bare component. If it is taken into account, the formation times 
of the aforementioned (anti-)collinear modes hence decrease by several orders of magnitude. As a consequence, then the contributions from the processes 
of quark/antiquark Bremsstrahlung and quark/antiquark pair annihilation into a single photon decrease much faster with increasing $\tau$.

Finally, we have compared our results to leading-order thermal photon production yields. We have seen that if one takes into account the full thermal 
mass of the quarks and antiquarks the photon numbers arising from leading-order thermal photon emission clearly outshine those from first-order 
non-equilibrium photon emission for photon energies of $\omega_{\vec{k}}=1-5$ GeV. On the other hand, our investigations indicate that first-order photon 
production in turn dominates for $\omega_{\vec{k}}\gtrsim5$ GeV. This does, however, not affect the quantitative accuracy of the recipe to address direct photon 
emission from the QGP phase by thermal calculations since both the thermal and the non-equilibrium contributions from this phase are outshone by direct 
photon emission arising from initial nucleon-nucleon scatterings and jet-medium interaction in that domain.

In summary, we have seen that our approach, which considers the photon number density for free asymptotic states, leads to 
physically reasonable results (no vacuum contribution, UV integrability) for this quantity. This is the case even though our ansatz for the 
time evolution of the QGP during a heavy-ion collision formally violates the Ward-Takahashi identities for the photon 
self-energy. The principal reason is that we (strictly speaking) make \textit{ad hoc} assumptions on the two-time dependence of the latter quantity 
by introducing time dependent quark/antiquark occupation numbers. On the other hand, it has been pointed out in \cite{Arleo:2004gn} that 
the conservation of QED gauge invariance and hence of the Ward-Takahashi identities remains challenging even if one tries to calculate the photon 
self-energy in a self-consistent framework such as the 2PI approach, where such assumptions are absent. A similar problem usually occurs when 
trying to calculate direct photon production within a transport framework: It has been shown in \cite{Knoll:1988dn} that the Thomas-Reiche-Kuhn 
sum rules, which are a direct consequence of gauge invariance of QED, impose restrictions on the actual applicability of the transport approaches on photon 
production from non-equilibrated hot hadronic matter presented in \cite{Nifenecker:1985,Ko:1985hq,Nakayama:1986zz,Bauer:1986zz,Neuhauser:1987}.

For our future investigations, however, the actual role of the Ward-Takahashi identities still requires further consideration. 
We have seen that even though they are formally violated within our model approach, this approach nevertheless leads to physically reasonable results for the 
(asymptotic) photon number density. In first sight, this seems to disprove our earlier conjecture that the artifacts encountered in 
\cite{Wang:2000pv,Wang:2001xh,Boyanovsky:2003qm} and still partly in \cite{Michler:2009hi} result from a violation of the Ward-Takahashi identities. 
Here one has to keep in mind, however, that these identities can be violated in two different ways:
\begin{itemize}
 \item Firstly, they can be violated directly by making \emph{ad hoc} assumptions on the two-time dependence of the photon-self energy. 
       This has been the case in \cite{Michler:2009hi} by introducing time-dependent occupation numbers in the one-loop thermal photon self-energy, 
       which on its own fulfills the Ward-Takahashi identities.
 \item On the other hand, they can also be violated indirectly by considering the `photon number density' at finite times and using 
       an inadequate definition of this quantity. This has been the case in \cite{Wang:2000pv,Wang:2001xh,Boyanovsky:2003qm}. The reason 
       is that the definition of the photon number density considered therein would only allow for an accordant interpretation if the 
       electromagnetic interaction was switched off at the point of time, $t$, at which said quantity is considered. By means of such a 
       switching, however, an effective violation of the Ward-Takahashi identities, which otherwise would be fulfilled, reoccurs. In this 
       context it is important to point out that an adequate definition of a transient particle number density is generally impossible 
       altogether for fundamental reasons except in some special settings \cite{Ruijsenaars:1976js,Fierz:1979ys,Thaller:1992ji,Scharf:1996zi}.
\end{itemize}
In particular, within the scope of our model description such an indirect violation would occur in addition to the direct one if we considered 
the photon number density at finite times. Consequently, it is of particular interest whether possibly only this indirect violation leads to 
artificial results.

\section*{Acknowledgments}
F. M. gratefully acknowledges financial support by the Helmholtz
Research School for Quark Matter Studies (H-QM) and by the Helmholtz
Graduate School for Hadron and Ion Research (HGS-HIRe for FAIR). This work has 
been supported (financially) by the Helmholtz International Center for FAIR within 
the framework of the LOEWE program (Landesoffensive zur Entwicklung 
Wissenschaftlich-\"Okonomischer Exzellenz) launched by the State of Hesse and by the 
German Federal Ministry of Education and Research (BMBF F\"orderkennzeichen 05P12RFCTB). 
The authors thank E. Fraga, S.\ Leupold, B.\ Schenke, J.\ Knoll, P.\ Danielewicz, and 
B.\ M\"uller for fruitful discussions.

\appendix

\section{Representation of the photon number density as an absolute square}
\label{sec:appa}
In this appendix, we show that each of the contributions (\ref{eq:2:photnum_split_bst})-(\ref{eq:2:photnum_split_pac}) can be written as the 
absolute square of a first-order QED transition amplitude and thus is positive (semi-)definite. For this purpose, we first undo the 
contraction of the individual contributions to the photon self-energy with $\gamma^{\mu\nu}(\vec{k})$. Then 
(\ref{eq:2:photnum_split_bst})-(\ref{eq:2:photnum_split_pac}) turn into
\begin{subequations}
 \label{eq:a:photnum_split}
 \begin{eqnarray}
  \left.2\omega_{\vec{k}}\frac{\dd^{6}n_{\gamma}(t)}{\dd^{3}x\dd^{3}k}\right|_{\text{BST}} 
    & = & \frac{\gamma^{\mu\nu}(\vec{k})}{(2\pi)^{3}}\int_{-\infty}^{t}\dd t_{1}\int_{-\infty}^{t}\dd t_{2}
          \ii\Pi^{\text{BST}}_{\nu\mu}(\vec{k},t_{1},t_{2})\ee^{\ii\omega_{\vec{k}}(t_{1}-t_{2})} \label{eq:a:photnum_split_bst} \ , \\
  \left.2\omega_{\vec{k}}\frac{\dd^{6}n_{\gamma}(t)}{\dd^{3}x\dd^{3}k}\right|_{\text{ANH}} 
    & = & \frac{\gamma^{\mu\nu}(\vec{k})}{(2\pi)^{3}}\int_{-\infty}^{t}\dd t_{1}\int_{-\infty}^{t}\dd t_{2}
          \ii\Pi^{\text{ANH}}_{\nu\mu}(\vec{k},t_{1},t_{2})\ee^{\ii\omega_{\vec{k}}(t_{1}-t_{2})} \label{eq:a:photnum_split_anh} \ , \\
  \left.2\omega_{\vec{k}}\frac{\dd^{6}n_{\gamma}(t)}{\dd^{3}x\dd^{3}k}\right|_{\text{PAC}} 
    & = & \frac{\gamma^{\mu\nu}(\vec{k})}{(2\pi)^{3}}\int_{-\infty}^{t}\dd t_{1}\int_{-\infty}^{t}\dd t_{2}
          \ii\Pi^{\text{PAC}}_{\nu\mu}(\vec{k},t_{1},t_{2})\ee^{\ii\omega_{\vec{k}}(t_{1}-t_{2})} \label{eq:a:photnum_split_pac} \ .
 \end{eqnarray}
\end{subequations}
It follows from (\ref{eq:2:prop_split_a})-(\ref{eq:2:prop_split_d}) that the contributions to the uncontracted photon self-energy read
\begin{subequations}
 \label{eq:a:pse_loop}
  \begin{align}
   \ii\Pi^{\text{BST}}_{\mu\nu}(\vec{k},t_{1},t_{2}) = 2e^{2}\int\frac{\dd^{3}p}{(2\pi)^{3}}
                                                        & \mbox{Tr}
                                                          \left\lbrace
                                                           \gamma_{\mu}\frac{\slashed{q}+m}{2q_{0}}
                                                           \gamma_{\nu}\frac{\slashed{p}+m}{2p_{0}}
                                                          \right\rbrace
                                                          n_{\text{F}}(q_{0})\left[1-n_{\text{F}}(p_{0})\right] \nonumber \\
                                                 \times & \ee^{-\ii(q_{0}-p_{0})(t_{1}-t_{2})} \label{eq:a:pse_loop_bst} \ , \\
   \ii\Pi^{\text{ANH}}_{\mu\nu}(\vec{k},t_{1},t_{2}) = e^{2}\int\frac{\dd^{3}p}{(2\pi)^{3}}
                                                        & \mbox{Tr}
                                                          \left\lbrace
                                                           \gamma_{\mu}\frac{\slashed{q}+m}{2q_{0}}
                                                           \gamma_{\nu}\frac{\slashed{\bar{p}}-m}{2p_{0}}
                                                          \right\rbrace
                                                          n_{\text{F}}(q_{0})n_{\text{F}}(p_{0}) \nonumber \\
                                                 \times & \ee^{-\ii(q_{0}+p_{0})(t_{1}-t_{2})} \label{eq:a:pse_loop_anh} \ , \\
   \ii\Pi^{\text{PAC}}_{\mu\nu}(\vec{k},t_{1},t_{2}) = e^{2}\int\frac{\dd^{3}p}{(2\pi)^{3}}
                                                        & \mbox{Tr}
                                                          \left\lbrace
                                                           \gamma_{\mu}\frac{\slashed{\bar{q}}-m}{2q_{0}}
                                                           \gamma_{\nu}\frac{\slashed{p}+m}{2p_{0}}
                                                          \right\rbrace
                                                          \left[1-n_{\text{F}}(q_{0})\right]\left[1-n_{\text{F}}(p_{0})\right] \nonumber \\
                                                 \times & \ee^{\ii(q_{0}+p_{0})(t_{1}-t_{2})} \label{eq:a:pse_loop_pac} \ .
  \end{align}
\end{subequations}
When incorporating the time evolution of the QGP into (\ref{eq:a:pse_loop_bst})-(\ref{eq:a:pse_loop_pac}) according to (\ref{eq:2:evol}) and 
(\ref{eq:2:evol_part})-(\ref{eq:2:evol_hole}), these expressions turn into
\begin{subequations}
 \label{eq:a:pse_loop_re}
  \begin{align}
   \ii\Pi^{\text{BST}}_{\mu\nu}(\vec{k},t_{1},t_{2}) = 2e^{2}\int\frac{\dd^{3}p}{(2\pi)^{3}}
                                                        & \mbox{Tr}
                                                          \left\lbrace
                                                           \gamma_{\mu}\frac{\slashed{q}+m}{2q_{0}}
                                                           \gamma_{\nu}\frac{\slashed{p}+m}{2p_{0}}
                                                          \right\rbrace
                                                          f_{\text{BST}}(q_{0},p_{0},t_{1})
                                                          f_{\text{BST}}(q_{0},p_{0},t_{2}) \nonumber \\
                                                 \times & \ee^{-\ii(q_{0}-p_{0})(t_{1}-t_{2})} 
                                                          \label{eq:a:pse_loop_re_bst} \ , \\
   \ii\Pi^{\text{ANH}}_{\mu\nu}(\vec{k},t_{1},t_{2}) = e^{2}\int\frac{\dd^{3}p}{(2\pi)^{3}}
                                                        & \mbox{Tr}
                                                          \left\lbrace
                                                           \gamma_{\mu}\frac{\slashed{q}+m}{2q_{0}}
                                                           \gamma_{\nu}\frac{\slashed{\bar{p}}-m}{2p_{0}}
                                                          \right\rbrace
                                                          f_{\text{ANH}}(q_{0},p_{0},t_{1})
                                                          f_{\text{ANH}}(q_{0},p_{0},t_{2}) \nonumber \\
                                                 \times & \ee^{-\ii(q_{0}+p_{0})(t_{1}-t_{2})}
                                                          \label{eq:a:pse_loop_re_anh} \ , \\
   \ii\Pi^{\text{PAC}}_{\mu\nu}(\vec{k},t_{1},t_{2}) = e^{2}\int\frac{\dd^{3}p}{(2\pi)^{3}}
                                                        & \mbox{Tr}
                                                          \left\lbrace
                                                           \gamma_{\mu}\frac{\slashed{\bar{q}}-m}{2q_{0}}
                                                           \gamma_{\nu}\frac{\slashed{p}+m}{2p_{0}}
                                                          \right\rbrace
                                                          f_{\text{PAC}}(q_{0},p_{0},t_{1})
                                                          f_{\text{PAC}}(q_{0},p_{0},t_{2}) \nonumber \\
                                                 \times & \ee^{\ii(q_{0}+p_{0})(t_{1}-t_{2})}
                                                          \label{eq:a:pse_loop_re_pac} \ .
  \end{align}
\end{subequations}
In order to keep the notation short, we have introduced
\begin{subequations}
 \begin{eqnarray}
  f_{\text{BST}}(q_{0},p_{0},t) & = & \sqrt{n_{\text{F}}(q_{0},t)\left[1-n_{\text{F}}(p_{0},t)\right]} \ , \\
  f_{\text{ANH}}(q_{0},p_{0},t) & = & \sqrt{n_{\text{F}}(q_{0},t)n_{\text{F}}(p_{0},t)} \ , \\
  f_{\text{PAC}}(q_{0},p_{0},t) & = & \sqrt{\left[1-n_{\text{F}}(q_{0},t)\right]\left[1-n_{\text{F}}(p_{0},t)\right]} \ .
 \end{eqnarray}
\end{subequations}
As the next step, we take into account that
\begin{subequations}
 \label{eq:spinor}
 \begin{eqnarray}
  \sum_{s}u(\vec{p},s)\bar{u}(\vec{p},s) = \frac{\slashed{p}+m}{2p_{0}} \label{eq:spinor_pos} \ , \\
  \sum_{s}v(\vec{p},s)\bar{v}(\vec{p},s) = \frac{\slashed{\bar{p}}-m}{2p_{0}} \label{eq:spinor_neg} \ .
 \end{eqnarray}
\end{subequations}
With the help of these relations, (\ref{eq:a:pse_loop_re_bst})-(\ref{eq:a:pse_loop_re_pac}) can be further rewritten as
\begin{subequations}
  \label{eq:a:pse_loop_decom}
  \begin{align}
   \ii\Pi^{\text{BST}}_{\mu\nu}(\vec{k},t_{1},t_{2}) = 2e^{2}\sum_{r,s}\int\frac{\dd^{3}p}{(2\pi)^{3}}
                                                            & \left[
                                                               \bar{u}(\vec{p},r)\gamma_{\mu}u(\vec{q},s)
                                                              \right]\cdot
                                                              \left[
                                                               \bar{u}(\vec{q},s)\gamma_{\nu}u(\vec{p},r)
                                                              \right] \nonumber \\
                                                     \times & f_{\text{BST}}(q_{0},p_{0},t_{1})
                                                              f_{\text{BST}}(q_{0},p_{0},t_{2})
                                                              \ee^{-\ii(q_{0}-p_{0})(t_{1}-t_{2})} \label{eq:a:pse_loop_decom_bst} \ , \\
   \ii\Pi^{\text{ANH}}_{\mu\nu}(\vec{k},t_{1},t_{2}) = e^{2}\sum_{r,s}\int\frac{\dd^{3}p}{(2\pi)^{3}}
                                                            & \left[
                                                               \bar{v}(\vec{p},r)\gamma_{\mu}u(\vec{q},s)
                                                              \right]\cdot
                                                              \left[
                                                               \bar{u}(\vec{q},s)\gamma_{\nu}v(\vec{p},r)
                                                              \right] \nonumber \\
                                                     \times & f_{\text{ANH}}(q_{0},p_{0},t_{1})
                                                              f_{\text{ANH}}(q_{0},p_{0},t_{2})
                                                              \ee^{-\ii(q_{0}+p_{0})(t_{1}-t_{2})} \label{eq:a:pse_loop_decom_anh} \ , \\
   \ii\Pi^{\text{PAC}}_{\mu\nu}(\vec{k},t_{1},t_{2}) = e^{2}\sum_{r,s}\int\frac{\dd^{3}p}{(2\pi)^{3}}
                                                            & \left[
                                                               \bar{u}(\vec{p},r)\gamma_{\mu}v(\vec{q},s)
                                                              \right]\cdot
                                                              \left[
                                                               \bar{v}(\vec{q},s)\gamma_{\nu}u(\vec{p},r)
                                                              \right] \nonumber \\
                                                     \times & f_{\text{PAC}}(q_{0},p_{0},t_{1})
                                                              f_{\text{PAC}}(q_{0},p_{0},t_{2})
                                                              \ee^{\ii(q_{0}+p_{0})(t_{1}-t_{2})} \label{eq:a:pse_loop_decom_pac} \ .
  \end{align}
\end{subequations}
If we now insert (\ref{eq:a:pse_loop_decom_bst})-(\ref{eq:a:pse_loop_decom_pac}) into (\ref{eq:a:photnum_split_bst})-(\ref{eq:a:photnum_split_anh}) 
and make use of relation (\ref{eq:2:polten}) we can finally rewrite the individual contributions to the photon number density as
\begin{subequations}
 \label{eq:a:asqr}
 \begin{align}
  \left.2\omega_{\vec{k}}\frac{\dd^{6}n_{\gamma}(t)}{\dd^{3}x\dd^{3}k}\right|_{\text{BST}}
    = \frac{2e^{2}}{(2\pi)^3}\sum_{\lambda,r,s}
      & \int\frac{\dd^{3}p}{(2\pi)^{3}}\left|\epsilon^{\mu,*}(\vec{k},\lambda)\bar{u}(\vec{p},r)\gamma_{\mu}u(\vec{q},s)\right. \nonumber \\
      & \left.\times\int_{-\infty}^{t}\dd u\mbox{ }f_{\text{BST}}(q_{0},p_{0},u)\ee^{-\ii(q_{0}-p_{0}-\omega_{\vec{k}})u}\right|^{2} \label{eq:a:asqr_bst} \ , \\
  \left.2\omega_{\vec{k}}\frac{\dd^{6}n_{\gamma}(t)}{\dd^{3}x\dd^{3}k}\right|_{\text{ANH}}
    = \frac{e^{2}}{(2\pi)^3}\sum_{\lambda,r,s}
      & \int\frac{\dd^{3}p}{(2\pi)^{3}}\left|\epsilon^{\mu,*}(\vec{k},\lambda)\bar{v}(\vec{p},r)\gamma_{\mu}u(\vec{q},s)\right. \nonumber \\
      & \left.\times\int_{-\infty}^{t}\dd u\mbox{ }f_{\text{ANH}}(q_{0},p_{0},u)\ee^{-\ii(q_{0}+p_{0}-\omega_{\vec{k}})u}\right|^{2} \label{eq:a:asqr_anh} \ , \\
   \left.2\omega_{\vec{k}}\frac{\dd^{6}n_{\gamma}(t)}{\dd^{3}x\dd^{3}k}\right|_{\text{PAC}}
    = \frac{e^{2}}{(2\pi)^3}\sum_{\lambda,r,s}
      & \int\frac{\dd^{3}p}{(2\pi)^{3}}\left|\epsilon^{\mu,*}(\vec{k},\lambda)\bar{u}(\vec{p},r)\gamma_{\mu}v(\vec{q},s)\right. \nonumber \\
      & \left.\times\int_{-\infty}^{t}\dd u\mbox{ }f_{\text{PAC}}(q_{0},p_{0},u)\ee^{\ii(q_{0}+p_{0}+\omega_{\vec{k}})u}\right|^{2} \label{eq:a:asqr_pac} \ .
 \end{align}
\end{subequations}
This completes the proof that (\ref{eq:2:photnum_split_bst})-(\ref{eq:2:photnum_split_pac}) can be expressed as absolute squares. Furthermore, taking a 
closer look at the underlying spinor structures shows that (\ref{eq:a:asqr_bst})-(\ref{eq:a:asqr_pac}) can be interpreted as the corresponding first-order 
QED process.


\begin{thebibliography}{40}
\expandafter\ifx\csname natexlab\endcsname\relax\def\natexlab#1{#1}\fi
\expandafter\ifx\csname bibnamefont\endcsname\relax
  \def\bibnamefont#1{#1}\fi
\expandafter\ifx\csname bibfnamefont\endcsname\relax
  \def\bibfnamefont#1{#1}\fi
\expandafter\ifx\csname citenamefont\endcsname\relax
  \def\citenamefont#1{#1}\fi
\expandafter\ifx\csname url\endcsname\relax
  \def\url#1{\texttt{#1}}\fi
\expandafter\ifx\csname urlprefix\endcsname\relax\def\urlprefix{URL }\fi
\providecommand{\bibinfo}[2]{#2}
\providecommand{\eprint}[2][]{\url{#2}}

\bibitem[{\citenamefont{Shuryak}(1978{\natexlab{a}})}]{Shuryak:1978ij}
\bibinfo{author}{\bibfnamefont{E.~V.} \bibnamefont{Shuryak}},
  \bibinfo{journal}{Phys. Lett. B} \textbf{\bibinfo{volume}{78}},
  \bibinfo{pages}{150} (\bibinfo{year}{1978}{\natexlab{a}}).

\bibitem[{\citenamefont{Shuryak}(1978{\natexlab{b}})}]{Shuryak:1977ut}
\bibinfo{author}{\bibfnamefont{E.~V.} \bibnamefont{Shuryak}},
  \bibinfo{journal}{Sov. Phys. JETP} \textbf{\bibinfo{volume}{47}},
  \bibinfo{pages}{212} (\bibinfo{year}{1978}{\natexlab{b}}).

\bibitem[{\citenamefont{Yagi et~al.}(2005)\citenamefont{Yagi, Hatsuda, and
  Miake}}]{Yag:2005}
\bibinfo{author}{\bibfnamefont{K.}~\bibnamefont{Yagi}},
  \bibinfo{author}{\bibfnamefont{T.}~\bibnamefont{Hatsuda}}, \bibnamefont{and}
  \bibinfo{author}{\bibfnamefont{Y.}~\bibnamefont{Miake}},
  \bibinfo{journal}{Camb. Monogr. Part. Phys. Nucl. Phys. Cosmol.}
  \textbf{\bibinfo{volume}{23}}, \bibinfo{pages}{1} (\bibinfo{year}{2005}).

\bibitem[{\citenamefont{M{\"u}ller and Nagle}(2006)}]{Muller:2006ee}
\bibinfo{author}{\bibfnamefont{B.}~\bibnamefont{M{\"u}ller}} \bibnamefont{and}
  \bibinfo{author}{\bibfnamefont{J.~L.} \bibnamefont{Nagle}},
  \bibinfo{journal}{Ann. Rev. Nucl. Part. Sci.} \textbf{\bibinfo{volume}{56}},
  \bibinfo{pages}{93} (\bibinfo{year}{2006}).

\bibitem[{\citenamefont{Friman et~al.}(2011)}]{Friman:2011zz}
\bibinfo{author}{\bibfnamefont{B.}~\bibnamefont{Friman}} \bibnamefont{et~al.},
  \bibinfo{journal}{Lect. Notes Phys.} \textbf{\bibinfo{volume}{814}},
  \bibinfo{pages}{1} (\bibinfo{year}{2011}).

\bibitem[{\citenamefont{Wang and Boyanovsky}(2001)}]{Wang:2000pv}
\bibinfo{author}{\bibfnamefont{S.-Y.} \bibnamefont{Wang}} \bibnamefont{and}
  \bibinfo{author}{\bibfnamefont{D.}~\bibnamefont{Boyanovsky}},
  \bibinfo{journal}{Phys. Rev. D} \textbf{\bibinfo{volume}{63}},
  \bibinfo{pages}{051702} (\bibinfo{year}{2001}).

\bibitem[{\citenamefont{Wang et~al.}(2002)\citenamefont{Wang, Boyanovsky, and
  Ng}}]{Wang:2001xh}
\bibinfo{author}{\bibfnamefont{S.-Y.} \bibnamefont{Wang}},
  \bibinfo{author}{\bibfnamefont{D.}~\bibnamefont{Boyanovsky}},
  \bibnamefont{and} \bibinfo{author}{\bibfnamefont{K.-W.} \bibnamefont{Ng}},
  \bibinfo{journal}{Nucl. Phys. A} \textbf{\bibinfo{volume}{699}},
  \bibinfo{pages}{819} (\bibinfo{year}{2002}).

\bibitem[{\citenamefont{Boyanovsky and de~Vega}(2003)}]{Boyanovsky:2003qm}
\bibinfo{author}{\bibfnamefont{D.}~\bibnamefont{Boyanovsky}} \bibnamefont{and}
  \bibinfo{author}{\bibfnamefont{H.~J.} \bibnamefont{de~Vega}},
  \bibinfo{journal}{Phys. Rev. D} \textbf{\bibinfo{volume}{68}},
  \bibinfo{pages}{065018} (\bibinfo{year}{2003}).

\bibitem[{\citenamefont{Michler et~al.}(2010)\citenamefont{Michler, Schenke,
  and Greiner}}]{Michler:2009hi}
\bibinfo{author}{\bibfnamefont{F.}~\bibnamefont{Michler}},
  \bibinfo{author}{\bibfnamefont{B.}~\bibnamefont{Schenke}}, \bibnamefont{and}
  \bibinfo{author}{\bibfnamefont{C.}~\bibnamefont{Greiner}},
  \bibinfo{journal}{Proceedings of the XLVII International Winter Meeting on
  Nuclear Physics}  (\bibinfo{year}{2010}), \eprint{arXiv: 0906.1734 [hep-ph]}.

\bibitem[{\citenamefont{Michler et~al.}(2013)\citenamefont{Michler, van Hees,
  Dietrich, Leupold, and Greiner}}]{Michler:2012mg}
\bibinfo{author}{\bibfnamefont{F.}~\bibnamefont{Michler}},
  \bibinfo{author}{\bibfnamefont{H.}~\bibnamefont{van Hees}},
  \bibinfo{author}{\bibfnamefont{D.~D.} \bibnamefont{Dietrich}},
  \bibinfo{author}{\bibfnamefont{S.}~\bibnamefont{Leupold}}, \bibnamefont{and}
  \bibinfo{author}{\bibfnamefont{C.}~\bibnamefont{Greiner}},
  \bibinfo{journal}{Annals Phys.} \textbf{\bibinfo{volume}{336}},
  \bibinfo{pages}{331} (\bibinfo{year}{2013}), \eprint{arXiv: 1208.6565
  [nucl-th]}.

\bibitem[{\citenamefont{Blaschke et~al.}(2010)\citenamefont{Blaschke, Schmidt,
  Smolyansky, and Tarakanov}}]{Blaschke:2009uy}
\bibinfo{author}{\bibfnamefont{D.}~\bibnamefont{Blaschke}},
  \bibinfo{author}{\bibfnamefont{S.}~\bibnamefont{Schmidt}},
  \bibinfo{author}{\bibfnamefont{S.}~\bibnamefont{Smolyansky}},
  \bibnamefont{and}
  \bibinfo{author}{\bibfnamefont{A.}~\bibnamefont{Tarakanov}},
  \bibinfo{journal}{Phys. Part. Nucl.} \textbf{\bibinfo{volume}{41}},
  \bibinfo{pages}{1004} (\bibinfo{year}{2010}).

\bibitem[{\citenamefont{Blaschke
  et~al.}(2011{\natexlab{a}})\citenamefont{Blaschke, Ropke, Schmidt,
  Smolyansky, and Tarakanov}}]{Blaschke:2010vs}
\bibinfo{author}{\bibfnamefont{D.}~\bibnamefont{Blaschke}},
  \bibinfo{author}{\bibfnamefont{G.}~\bibnamefont{Ropke}},
  \bibinfo{author}{\bibfnamefont{S.}~\bibnamefont{Schmidt}},
  \bibinfo{author}{\bibfnamefont{S.}~\bibnamefont{Smolyansky}},
  \bibnamefont{and}
  \bibinfo{author}{\bibfnamefont{A.}~\bibnamefont{Tarakanov}},
  \bibinfo{journal}{Contrib. Plasma Phys.} \textbf{\bibinfo{volume}{51}},
  \bibinfo{pages}{451} (\bibinfo{year}{2011}{\natexlab{a}}).

\bibitem[{\citenamefont{Smolyansky et~al.}(2010)\citenamefont{Smolyansky,
  Blaschke, Chertilin, Roepke, and Tarakanov}}]{Smolyansky:2010as}
\bibinfo{author}{\bibfnamefont{S.}~\bibnamefont{Smolyansky}},
  \bibinfo{author}{\bibfnamefont{D.}~\bibnamefont{Blaschke}},
  \bibinfo{author}{\bibfnamefont{A.}~\bibnamefont{Chertilin}},
  \bibinfo{author}{\bibfnamefont{G.}~\bibnamefont{Roepke}}, \bibnamefont{and}
  \bibinfo{author}{\bibfnamefont{A.}~\bibnamefont{Tarakanov}}
  (\bibinfo{year}{2010}), \eprint{arXiv:1012.0559 [physics.plasma-ph]}.

\bibitem[{\citenamefont{Blaschke
  et~al.}(2011{\natexlab{b}})\citenamefont{Blaschke, Ropke, Dmitriev,
  Smolyansky, and Tarakanov}}]{Blaschke:2011af}
\bibinfo{author}{\bibfnamefont{D.}~\bibnamefont{Blaschke}},
  \bibinfo{author}{\bibfnamefont{G.}~\bibnamefont{Ropke}},
  \bibinfo{author}{\bibfnamefont{V.}~\bibnamefont{Dmitriev}},
  \bibinfo{author}{\bibfnamefont{S.}~\bibnamefont{Smolyansky}},
  \bibnamefont{and} \bibinfo{author}{\bibfnamefont{A.}~\bibnamefont{Tarakanov}}
  (\bibinfo{year}{2011}{\natexlab{b}}), \eprint{arXiv: 1101.6021
  [physics.plasma-ph]}.

\bibitem[{\citenamefont{Blaschke
  et~al.}(2011{\natexlab{c}})\citenamefont{Blaschke, Dmitriev, Ropke, and
  Smolyansky}}]{Blaschke:2011is}
\bibinfo{author}{\bibfnamefont{D.}~\bibnamefont{Blaschke}},
  \bibinfo{author}{\bibfnamefont{V.}~\bibnamefont{Dmitriev}},
  \bibinfo{author}{\bibfnamefont{G.}~\bibnamefont{Ropke}}, \bibnamefont{and}
  \bibinfo{author}{\bibfnamefont{S.}~\bibnamefont{Smolyansky}},
  \bibinfo{journal}{Phys. Rev. D} \textbf{\bibinfo{volume}{84}},
  \bibinfo{pages}{085028} (\bibinfo{year}{2011}{\natexlab{c}}).

\bibitem[{\citenamefont{Bethe and Heitler}(1934)}]{Bethe01081934}
\bibinfo{author}{\bibfnamefont{H.}~\bibnamefont{Bethe}} \bibnamefont{and}
  \bibinfo{author}{\bibfnamefont{W.}~\bibnamefont{Heitler}},
  \bibinfo{journal}{Proceedings of the Royal Society of London. Series A}
  \textbf{\bibinfo{volume}{146}}, \bibinfo{pages}{83} (\bibinfo{year}{1934}).

\bibitem[{\citenamefont{Kluger et~al.}(1992)\citenamefont{Kluger, Eisenberg,
  Svetitsky, Cooper, and Mottola}}]{Kluger:1992gb}
\bibinfo{author}{\bibfnamefont{Y.}~\bibnamefont{Kluger}},
  \bibinfo{author}{\bibfnamefont{J.}~\bibnamefont{Eisenberg}},
  \bibinfo{author}{\bibfnamefont{B.}~\bibnamefont{Svetitsky}},
  \bibinfo{author}{\bibfnamefont{F.}~\bibnamefont{Cooper}}, \bibnamefont{and}
  \bibinfo{author}{\bibfnamefont{E.}~\bibnamefont{Mottola}},
  \bibinfo{journal}{Phys. Rev. D} \textbf{\bibinfo{volume}{45}},
  \bibinfo{pages}{4659} (\bibinfo{year}{1992}).

\bibitem[{\citenamefont{Schmidt et~al.}(1998)\citenamefont{Schmidt, Blaschke,
  Ropke, Smolyansky, Prozorkevich et~al.}}]{Schmidt:1998vi}
\bibinfo{author}{\bibfnamefont{S.}~\bibnamefont{Schmidt}},
  \bibinfo{author}{\bibfnamefont{D.}~\bibnamefont{Blaschke}},
  \bibinfo{author}{\bibfnamefont{G.}~\bibnamefont{Ropke}},
  \bibinfo{author}{\bibfnamefont{S.}~\bibnamefont{Smolyansky}},
  \bibinfo{author}{\bibfnamefont{A.}~\bibnamefont{Prozorkevich}},
  \bibnamefont{et~al.}, \bibinfo{journal}{Int. J. Mod. Phys. E}
  \textbf{\bibinfo{volume}{7}}, \bibinfo{pages}{709} (\bibinfo{year}{1998}).

\bibitem[{\citenamefont{Pervushin and Skokov}(2006)}]{Pervushin:2006vh}
\bibinfo{author}{\bibfnamefont{V.}~\bibnamefont{Pervushin}} \bibnamefont{and}
  \bibinfo{author}{\bibfnamefont{V.~V.} \bibnamefont{Skokov}},
  \bibinfo{journal}{Acta Phys. Polon. B} \textbf{\bibinfo{volume}{37}},
  \bibinfo{pages}{2587} (\bibinfo{year}{2006}).

\bibitem[{\citenamefont{Blaschke et~al.}(2013)\citenamefont{Blaschke, Kaempfer,
  Schmidt, Panferov, Prozorkevich et~al.}}]{Blaschke:2013ip}
\bibinfo{author}{\bibfnamefont{D.}~\bibnamefont{Blaschke}},
  \bibinfo{author}{\bibfnamefont{B.}~\bibnamefont{Kaempfer}},
  \bibinfo{author}{\bibfnamefont{S.}~\bibnamefont{Schmidt}},
  \bibinfo{author}{\bibfnamefont{A.}~\bibnamefont{Panferov}},
  \bibinfo{author}{\bibfnamefont{A.}~\bibnamefont{Prozorkevich}},
  \bibnamefont{et~al.}, \bibinfo{journal}{Phys. Rev. D}
  \textbf{\bibinfo{volume}{88}}, \bibinfo{pages}{045017}
  (\bibinfo{year}{2013}).

\bibitem[{\citenamefont{Serreau}(2004)}]{Serreau:2003wr}
\bibinfo{author}{\bibfnamefont{J.}~\bibnamefont{Serreau}},
  \bibinfo{journal}{JHEP} \textbf{\bibinfo{volume}{05}}, \bibinfo{pages}{078}
  (\bibinfo{year}{2004}).

\bibitem[{\citenamefont{Heinz and Kolb}(2002)}]{Heinz:2001xi}
\bibinfo{author}{\bibfnamefont{U.~W.} \bibnamefont{Heinz}} \bibnamefont{and}
  \bibinfo{author}{\bibfnamefont{P.~F.} \bibnamefont{Kolb}},
  \bibinfo{journal}{Nucl. Phys. A} \textbf{\bibinfo{volume}{702}},
  \bibinfo{pages}{269} (\bibinfo{year}{2002}).

\bibitem[{\citenamefont{Arnold et~al.}(2001)\citenamefont{Arnold, Moore, and
  Yaffe}}]{Arnold:2001ms}
\bibinfo{author}{\bibfnamefont{P.~B.} \bibnamefont{Arnold}},
  \bibinfo{author}{\bibfnamefont{G.~D.} \bibnamefont{Moore}}, \bibnamefont{and}
  \bibinfo{author}{\bibfnamefont{L.~G.} \bibnamefont{Yaffe}},
  \bibinfo{journal}{JHEP} \textbf{\bibinfo{volume}{0112}}, \bibinfo{pages}{009}
  (\bibinfo{year}{2001}).

\bibitem[{\citenamefont{van Hees et~al.}(2011)\citenamefont{van Hees, Gale, and
  Rapp}}]{vanHees:2011vb}
\bibinfo{author}{\bibfnamefont{H.}~\bibnamefont{van Hees}},
  \bibinfo{author}{\bibfnamefont{C.}~\bibnamefont{Gale}}, \bibnamefont{and}
  \bibinfo{author}{\bibfnamefont{R.}~\bibnamefont{Rapp}},
  \bibinfo{journal}{Phys. Rev. C} \textbf{\bibinfo{volume}{84}},
  \bibinfo{pages}{054906} (\bibinfo{year}{2011}).

\bibitem[{\citenamefont{Turbide et~al.}(2004)\citenamefont{Turbide, Rapp, and
  Gale}}]{Turbide:2003si}
\bibinfo{author}{\bibfnamefont{S.}~\bibnamefont{Turbide}},
  \bibinfo{author}{\bibfnamefont{R.}~\bibnamefont{Rapp}}, \bibnamefont{and}
  \bibinfo{author}{\bibfnamefont{C.}~\bibnamefont{Gale}},
  \bibinfo{journal}{Phys. Rev. C} \textbf{\bibinfo{volume}{69}},
  \bibinfo{pages}{014903} (\bibinfo{year}{2004}).

\bibitem[{\citenamefont{Turbide et~al.}(2005)\citenamefont{Turbide, Gale, Jeon,
  and Moore}}]{Turbide:2005fk}
\bibinfo{author}{\bibfnamefont{S.}~\bibnamefont{Turbide}},
  \bibinfo{author}{\bibfnamefont{C.}~\bibnamefont{Gale}},
  \bibinfo{author}{\bibfnamefont{S.}~\bibnamefont{Jeon}}, \bibnamefont{and}
  \bibinfo{author}{\bibfnamefont{G.~D.} \bibnamefont{Moore}},
  \bibinfo{journal}{Phys. Rev. C} \textbf{\bibinfo{volume}{72}},
  \bibinfo{pages}{014906} (\bibinfo{year}{2005}).

\bibitem[{\citenamefont{Gale}(2009)}]{Gale:2009gc}
\bibinfo{author}{\bibfnamefont{C.}~\bibnamefont{Gale}} (\bibinfo{year}{2009}),
  \eprint{arXiv:0904.2184 [hep-ph]}.

\bibitem[{\citenamefont{Fraga et~al.}(2005{\natexlab{a}})\citenamefont{Fraga,
  Gelis, and Schiff}}]{Fraga:2003sn}
\bibinfo{author}{\bibfnamefont{E.}~\bibnamefont{Fraga}},
  \bibinfo{author}{\bibfnamefont{F.}~\bibnamefont{Gelis}}, \bibnamefont{and}
  \bibinfo{author}{\bibfnamefont{D.}~\bibnamefont{Schiff}},
  \bibinfo{journal}{Phys. Rev. D} \textbf{\bibinfo{volume}{71}},
  \bibinfo{pages}{085015} (\bibinfo{year}{2005}{\natexlab{a}}).

\bibitem[{\citenamefont{Fraga et~al.}(2005{\natexlab{b}})\citenamefont{Fraga,
  Gelis, and Schiff}}]{Fraga:2004ur}
\bibinfo{author}{\bibfnamefont{E.~S.} \bibnamefont{Fraga}},
  \bibinfo{author}{\bibfnamefont{F.}~\bibnamefont{Gelis}}, \bibnamefont{and}
  \bibinfo{author}{\bibfnamefont{D.}~\bibnamefont{Schiff}},
  \bibinfo{journal}{AIP Conf. Proc.} \textbf{\bibinfo{volume}{739}},
  \bibinfo{pages}{437} (\bibinfo{year}{2005}{\natexlab{b}}).

\bibitem[{\citenamefont{Arleo et~al.}(2004)}]{Arleo:2004gn}
\bibinfo{author}{\bibfnamefont{F.}~\bibnamefont{Arleo}} \bibnamefont{et~al.}
  (\bibinfo{year}{2004}), \eprint{hep-ph/0311131}.

\bibitem[{\citenamefont{Knoll and Guet}(1989)}]{Knoll:1988dn}
\bibinfo{author}{\bibfnamefont{J.}~\bibnamefont{Knoll}} \bibnamefont{and}
  \bibinfo{author}{\bibfnamefont{C.}~\bibnamefont{Guet}},
  \bibinfo{journal}{{Nucl. Phys. A}} \textbf{\bibinfo{volume}{494}},
  \bibinfo{pages}{334} (\bibinfo{year}{1989}).

\bibitem[{\citenamefont{Nifenecker and Bondorf}(1985)}]{Nifenecker:1985}
\bibinfo{author}{\bibfnamefont{H.}~\bibnamefont{Nifenecker}} \bibnamefont{and}
  \bibinfo{author}{\bibfnamefont{J.}~\bibnamefont{Bondorf}},
  \bibinfo{journal}{{Nucl. Phys. A}} \textbf{\bibinfo{volume}{442}},
  \bibinfo{pages}{478} (\bibinfo{year}{1985}).

\bibitem[{\citenamefont{Ko et~al.}(1985)\citenamefont{Ko, Bertsch, and
  Aichelin}}]{Ko:1985hq}
\bibinfo{author}{\bibfnamefont{C.}~\bibnamefont{Ko}},
  \bibinfo{author}{\bibfnamefont{G.}~\bibnamefont{Bertsch}}, \bibnamefont{and}
  \bibinfo{author}{\bibfnamefont{J.}~\bibnamefont{Aichelin}},
  \bibinfo{journal}{Phys. Rev. C} \textbf{\bibinfo{volume}{31}},
  \bibinfo{pages}{2324} (\bibinfo{year}{1985}).

\bibitem[{\citenamefont{Nakayama and Bertsch}(1986)}]{Nakayama:1986zz}
\bibinfo{author}{\bibfnamefont{K.}~\bibnamefont{Nakayama}} \bibnamefont{and}
  \bibinfo{author}{\bibfnamefont{G.}~\bibnamefont{Bertsch}},
  \bibinfo{journal}{Phys. Rev. C} \textbf{\bibinfo{volume}{34}},
  \bibinfo{pages}{2190} (\bibinfo{year}{1986}).

\bibitem[{\citenamefont{Bauer et~al.}(1986)\citenamefont{Bauer, Bertsch,
  Cassing, and Mosel}}]{Bauer:1986zz}
\bibinfo{author}{\bibfnamefont{W.}~\bibnamefont{Bauer}},
  \bibinfo{author}{\bibfnamefont{G.}~\bibnamefont{Bertsch}},
  \bibinfo{author}{\bibfnamefont{W.}~\bibnamefont{Cassing}}, \bibnamefont{and}
  \bibinfo{author}{\bibfnamefont{U.}~\bibnamefont{Mosel}},
  \bibinfo{journal}{Phys. Rev. C} \textbf{\bibinfo{volume}{34}},
  \bibinfo{pages}{2127} (\bibinfo{year}{1986}).

\bibitem[{\citenamefont{Neuhauser and Koonin}(1987)}]{Neuhauser:1987}
\bibinfo{author}{\bibfnamefont{D.}~\bibnamefont{Neuhauser}} \bibnamefont{and}
  \bibinfo{author}{\bibfnamefont{S.}~\bibnamefont{Koonin}},
  \bibinfo{journal}{{Nucl. Phys. A}} \textbf{\bibinfo{volume}{462}},
  \bibinfo{pages}{163} (\bibinfo{year}{1987}).

\bibitem[{\citenamefont{Ruijsenaars}(1977)}]{Ruijsenaars:1976js}
\bibinfo{author}{\bibfnamefont{S.}~\bibnamefont{Ruijsenaars}},
  \bibinfo{journal}{J. Math. Phys.} \textbf{\bibinfo{volume}{18}},
  \bibinfo{pages}{720} (\bibinfo{year}{1977}).

\bibitem[{\citenamefont{Fierz and Scharf}(1979)}]{Fierz:1979ys}
\bibinfo{author}{\bibfnamefont{H.}~\bibnamefont{Fierz}} \bibnamefont{and}
  \bibinfo{author}{\bibfnamefont{G.}~\bibnamefont{Scharf}},
  \bibinfo{journal}{Helv. Phys. Acta} \textbf{\bibinfo{volume}{52}},
  \bibinfo{pages}{437} (\bibinfo{year}{1979}).

\bibitem[{\citenamefont{Thaller}(1992)}]{Thaller:1992ji}
\bibinfo{author}{\bibfnamefont{B.}~\bibnamefont{Thaller}},
  \emph{\bibinfo{title}{The Dirac Equation}}, Texts and monographs in physics
  (\bibinfo{publisher}{Springer}, \bibinfo{year}{1992}).

\bibitem[{\citenamefont{Scharf}(1995)}]{Scharf:1996zi}
\bibinfo{author}{\bibfnamefont{G.}~\bibnamefont{Scharf}},
  \emph{\bibinfo{title}{Finite quantum electrodynamics: The Causal approach}},
  Texts and monographs in physics (\bibinfo{publisher}{Springer},
  \bibinfo{year}{1995}).
\end{thebibliography}
\end{document}